\documentclass[journal]{IEEEtran}

\usepackage{cite}
\usepackage[pdftex]{graphicx}
\graphicspath{{images/}{logos/}}
\DeclareGraphicsExtensions{.pdf,.jpeg,.png}

\usepackage{amsmath}
\newcommand\norm[1]{\left\lVert#1\right\rVert}
\renewcommand\vec[1]{\mathbf{#1}}
\newcommand\Var[1]{\mathop{\textrm{Var}}\left[#1\right]}

\usepackage{amsfonts}
\usepackage{amssymb}

\usepackage{array}
\usepackage{url}
\usepackage[final,nomargin,inline,author=]{fixme}
\fxsetup{theme=color,mode=multiuser}
\definecolor{fxwarning}{rgb}{0.8,0.0000,0.0000}

\usepackage{siunitx}
\usepackage{textgreek}

\usepackage{tabularx}
\usepackage{adjustbox}

\newcommand*\around{{\raise.17ex\hbox{$\scriptstyle\mathtt{\sim}$}}}

\usepackage{booktabs}

\usepackage{algpseudocode}% http://ctan.org/pkg/algorithmicx

\usepackage{chemformula}

\usepackage{commath}

\usepackage{hyperref}
\hypersetup{
  pdftitle={Design Automation of Photonic Resonator Weights},
  pdfauthor={Thomas Ferreira de Lima}
}

\begin{document}
%% IEEE TEMPLATE
% paper title
% Titles are generally capitalized except for words such as a, an, and, as,
% at, but, by, for, in, nor, of, on, or, the, to and up, which are usually
% not capitalized unless they are the first or last word of the title.
% Linebreaks \\ can be used within to get better formatting as desired.
% Do not put math or special symbols in the title.
\title{Design Automation of Photonic Resonator Weights}
%
%
% author names and IEEE memberships
% note positions of commas and nonbreaking spaces ( ~ ) LaTeX will not break
% a structure at a ~ so this keeps an author's name from being broken across
% two lines.
% use \thanks{} to gain access to the first footnote area
% a separate \thanks must be used for each paragraph as LaTeX2e's \thanks
% was not built to handle multiple paragraphs
%

\author{%
        Thomas~Ferreira~de~Lima,
        Eli~A.~Doris\textsuperscript{*},
        Simon~Bilodeau,
        Weipeng~Zhang,
        Aashu~Jha,
        Hsuan-Tung~Peng,
        Eric~C.~Blow,
        Chaoran~Huang,
        Alexander~N.~Tait,
        Bhavin~J.~Shastri,%~\IEEEmembership{Senior~Member,~IEEE,}
        ~and Paul~R.~Prucnal%,~\IEEEmembership{Life~Fellow,~IEEE}
        %<-this % stops a space
  \thanks{\textsuperscript{*}Corresponding author: edoris@princeton.edu}%
  \thanks{T.F.L., E.A.D, S.B., W.Z., A.J, H.-T.P, E.C.B, C.H., A.N.T., B.J.S. and P.R.P. are with the Department of Electrical and Computer Engineering, Princeton University, Princeton, NJ 08544, USA.}%
  \thanks{T.F.L is currently with NEC Laboratories America, Princeton NJ, USA}%
  \thanks{C.H. is currently with the Department of Electrical Engineering, The Chinese University of Hong Kong, Hong Kong SAR, China.}%
  \thanks{A.T. is currently with the Department of Electrical and Computer Engineering, Queen's University, Kingston, ON K7L 3N6, Canada}
  \thanks{B.J.S. is currently with the Department of Physics, Engineering Physics \& Astronomy, Queen's University, Kingston, ON KL7 3N6, Canada.}%

  % \thanks{Manuscript received April 19, 2005; revised August 26, 2015.}
}

\maketitle

% As a general rule, do not put math, special symbols or citations
% in the abstract or keywords.
\begin{abstract}
% \abstract{
% \fxnote{148/200 words}
Neuromorphic photonic processors based on resonator weight banks are an emerging candidate technology for enabling modern artificial intelligence (AI) in high speed, analog systems. These purpose-built analog devices implement vector multiplications with the physics of resonator devices, offering efficiency, latency, and throughput advantages over equivalent electronic circuits. Along with these advantages, however, often comes the difficult challenges of compensation for fabrication variations and environmental disturbances.
In this paper we review sources of variation and disturbances from our experiments, as well as mathematically define quantities that model them. Then, we introduce how the physics of resonators can be exploited to weight and sum multiwavelength signals.
Finally, we outline automated design and control methodologies necessary to create practical, manufacturable, and high accuracy/precision resonator weight banks that can withstand operating conditions in the field. This represents a road map for unlocking the potential of resonator weight banks in practical deployment scenarios.
% }
\end{abstract}

% Note that keywords are not normally used for peerreview papers.

% \keywords{silicon photonics, RF photonics, programmable photonics}

% For peer review papers, you can put extra information on the cover
% page as needed:
% \ifCLASSOPTIONpeerreview
% \begin{center} \bfseries EDICS Category: 3-BBND \end{center}
% \fi
%
% For peerreview papers, this IEEEtran command inserts a page break and
% creates the second title. It will be ignored for other modes.
% \IEEEpeerreviewmaketitle

% \maketitle

\section{Introduction}

  Most of the recent successes in artificial intelligence (AI) are tied to the ability to perform an increasingly large amount of computation on a mathematical representation of information (i.e., data). In machine learning (ML) algorithms, data is first rearranged into mathematical objects called vectors living inside of a vector space. These vectors can then be manipulated by linear and nonlinear operations resulting in another vector that can be translated into a useful result. For example, in an image classification task the input image is typically encoded into a vector containing three 2-dimensional arrays of numbers representing the intensity values of red, green, and blue for each pixel. A classifier then computes a function that maps the image vector into a classification vector, which is a 1-dimensional array with each component representing a human-interpretable label (e.g., cat, dog).
  In modern AI algorithms, especially deep neural networks, most of the hardware processing power in conventional processors is devoted to memory access for parallel linear operations on vectors~\cite{horowitz_computings_2014}, resulting in a limitation as models scaled up in size. Accordingly, specialized hardware that efficiently perform linear operations have been introduced to accommodate the increasing demand for AI. These include GPUs~\cite{raina_large-scale_2009}, TPU~\cite{jouppi_-datacenter_2017} and neuromorphic processors~\cite{davies_loihi_2018,akopyan_truenorth_2015}.
  % These innovative hardware solutions have enabled more ambitious AI models to exist than would be possible with traditional von Neumann computing architectures alone.

  In-memory computing and multi-processing techniques are two innovations that have allowed specialized hardware such as multi-core CPUs and GPUs to address the increasing demand for data processing. These techniques, however, were not a significant enough leap forward to close the gap between current computing capabilities and the needs of evolving AI tasks~\cite{Merolla2014,davies_loihi_2018}, especially for high-frequency computation with low latency. Brain-inspired, or neuromorphic, computing schemes have emerged as a potential alternative for addressing these deficiencies in power efficiency and speed.
  Neuromorphic processor architectures are optimized on a hardware level to run neural network based algorithms as fast and efficiently as possible~\cite{mead_neuromorphic_1990}.
  Such architectures derive these benefits from making use of analog systems in which the computation is inherently tied to the physics of the device itself, rather than defined in software on a generalized digital architecture. Analog systems, however, must be engineered to control noise in such a way that the signals of interest are not corrupted by the processing itself, a task which is nearly trivial in digital electronics. It is also crucial to design systems which are user-transparent and are capable of interfacing with existing computational systems.

  Most optical phenomena are linear, making multiplication and addition operations easy to implement using photonic devices.
  Photonic systems can therefore benefit from the body of knowledge in the field of ML, which has over the decades created algorithms that are heavily dominated by linear computation. These are directly utilized in photonic processors that operate using analog inputs and outputs.
  As an added benefit, information in different channels can be encoded onto non-interacting optical carriers using wavelength-division multiplexing (WDM) allowing for a high density of signals on a single waveguide. Utilizing WDM in turn allows for the use of resonator devices that produce controllable effects only around their specific resonant wavelength and do not interact with other optical carriers.

  In neuromorphic photonics \cite{prucnal_neuromorphic_2017}, the device responsible for linear computation is the weight bank, comprised of a series of resonators with unique resonant wavelengths, which selectively weights (multiply) a series of incoming WDM signals~\cite{Tait2016} and optically sums all of the resulting light. These have been optimized specifically for real-time weighted summation of high-bandwidth analog signals. The most common designs electronically control the weights and encode input signals onto multi-wavelength lightwaves. Summation using photodetectors returns an analog electrical signal, which can either be read out directly or used to modulate another optical carrier. This multivariate processing ability has proven to be particularly useful in applications with analog inputs and outputs, such as microwave signal processing, ultrafast robotic control \cite{ferreira_de_lima_machine_2019}, and neuromorphic computing \cite{shastri_photonics_2021,huang_silicon_2021}.
  
  One of the main engineering challenges facing weight banks is the ability to precisely and accurately control weight values, which is made difficult due to fabrication variations and environmental disturbances. In this paper, we aim to outline automated methodologies for designing and controlling resonator weights in order to enable practical WDM photonic processors. In sections \ref{overview} and \ref{resphys} we introduce the necessary background for understanding and engineering resonator weight banks. Next, in section \ref{sec:calibration and control}, we introduce a practical model of weight banks and a feedforward algorithm that can be used for calibration and control of real devices. Finally, in section \ref{design}, we discuss options for designs that allow for compensation of fabrication variations and environmental disturbances as well as a feedback algorithm that can be used if a weight bank is manufactured with built-in weight sensing. The aim is to allow readers to design systems utilizing resonator weight banks that are robust enough to be brought beyond laboratory benches and into the field.

\section{Weight Bank Overview \label{overview}}
    \subsection{Matrix-Vector Multiplication}
    Tensor operations, which are the core of many machine learning algorithms, can be implemented via a collection of matrix-vector multiplications. Matrix-vector multiplications, in turn, are typically decomposed into a number of vector-vector dot products where the first vector corresponds to the rows of the matrix and the second vector (the vector of the matrix-vector operation) is identical for all such sub-operations. Hardware units for performing matrix-vector units can be modularized in the same way, with the most basic units implementing vector-vector dot products. Vector-vector dot products can be written mathematically as
    
    \begin{gather}
        y(t)=\sum_{i=1}^nw_i(t)x_i(t) \label{eq:vectordotproduct}
    \end{gather}
    
    where $\vec{x}=[x_1,\ldots,x_n]^T$ is a vector of incoming signal values, $\vec{w}=[w_1,\ldots,w_n]^T$ is a vector of weight values, and $y(t)$ is the output. Although $\vec{x}(t)$ and $\vec{w}(t)$ function identically mathematically, we make the distinction between signals and weights due to the distinct forms that these vectors take in photonic hardware. 
    
    Since this operation is a collection of multiplication operations with a summation, it is common to refer to the performance of a system in terms of multiply-accumulate (MAC) operations~\cite{frantz_digital_2000}. The number of MAC operations for a particular computation is hardware-agnostic, although some architectures, such as systolic arrays~\cite{ramacher_design_1991,jouppi_-datacenter_2017}, implement MACs directly while others, such as the WDM architecture analyzed in this paper, implement weights as individual units and perform all summation simultaneously.
    
    Many applications involve multiplying a batch of vectors with the same matrix, such as running inference tasks with input data on a fixed (pre-trained) neural network classifier.
    This approach optimizes inference tasks for high bandwidth signals and low latency result.
    In such cases, it can be advantageous to design analog MAC units where input signals $\vec{x}(t)$ can be modulated on fast time scales and are multiplied by ``fixed'' weights $\vec{w}$ that only change on relatively-slow time scales.
    
    \subsection{Role of Individual Resonators}
    
\begin{figure}[!t]
      \centering
      \includegraphics[width=\linewidth]{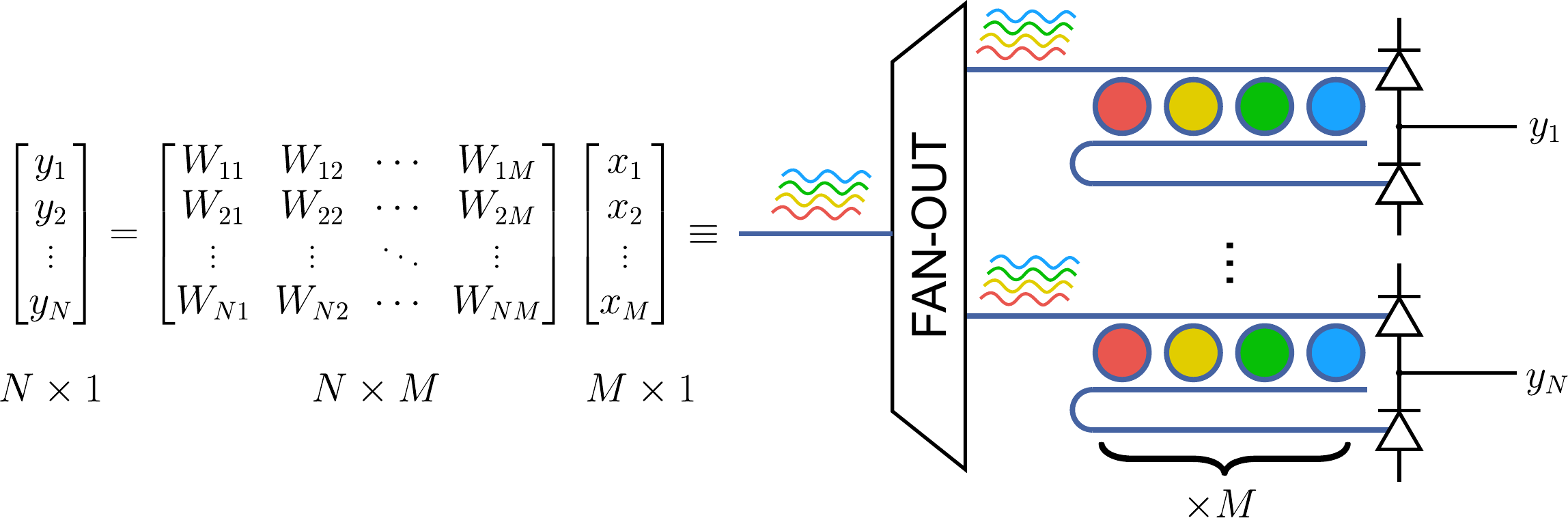}
      % \caption{Schematic of a WDM-based MVM. Each matrix row of size $N$ corresponds to a weight bank with $N$ resonator weights. At each point in time, the output of the weight bank corresponds to the vector dot product between a matrix row and the multiplicand vector. The higher the bandwidth of the signal $x(t)$, the more computations this system does per second. By stacking $M$ such weight banks, $M$ dot product operations can be completed in parallel without sacrifice to bandwidth.
      % }
      \caption[Schematic of WDM-based MVM.]{Schematic of WDM-based MVM. Each matrix row of size $N$ corresponds to a weight bank with $M$ resonator weights. At each point in time, the output of the weight bank corresponds to the vector dot product between a matrix row and the multiplicand vector. The higher the bandwidth of the signal $x(t)$, the more computations this system does per second. By stacking $N$ such weight banks, $N$ dot product operations can be completed in parallel without sacrifice to bandwidth.
      }
    \label{fig:MVM}
    \end{figure}

    Resonator-based weights are specifically designed to take advantage of wavelength-division multiplexing (WDM), in which separate signals are encoded on optical carriers with non-overlapping wavelengths. In optical communications, multigigahertz signals are modulated as amplitude or phase changes on a continuous-wave (CW) carrier traveling through a bus waveguide. With WDM, hundreds of CW carriers with unique wavelengths can coexist in a single waveguide without interfering with one another, effectively creating many independent information channels within a single physical channel. An array of resonators, usually microring resonators, with resonances matching the optical carrier wavelengths is used to access each channel independently. Each resonator, when in resonance, transfers all of the optical energy traveling on one waveguide onto another waveguide. This property is utilized, for example, to enable compact optical interconnections on photonic chips for use in telecommunication systems~\cite{xu_cascaded_2006}. If the resonances can be independently tuned, the amount of energy transferred between waveguides can be controlled for each channel, resulting in a precise weighting (multiplication) of the input WDM signal. Using this scheme for implementing matrix-vector multiplications is depicted in Fig.~\ref{fig:MVM}.

    Silicon photonics has emerged as a popular platform for creating photonic integrated circuits (PICs), owing to low loss and compatibility with commercial foundries~\cite{rahim_open-access_2018}. Some process design kits (PDKs) are being generated~\cite{Giewont2019,chrostowski_silicon_2019}, but optical devices are very sensitive to temperature and fabrication variations. This poses unique challenges to the design of PICs on top of those usually associated with analog devices. In the case of microring resonator filters, this sensitivity can be counteracted via calibration and control techniques~\cite{padmaraju_resolving_2014}. Early demonstrations used integrated sensors to coarsely lock microring resonators on the transmission wavelength~\cite{derose_silicon_2010}. These involved applying a compensating electronic signal (either current or voltage) based on a measured property, mathematical state model, or both. This can be achieved easily with microcontroller circuits placed in a CMOS electronic circuit nearby or monolithically integrated on the same die~\cite{grillanda_non-invasive_2014}. Resonator weights, on the other hand, require more sophisticated control schemes, because the resonator not only needs to be stabilized near resonance, but also need to be fine-tuned with a high degree of accuracy.

\section{Resonator Weight Physics \label{resphys}}

    \subsection{Resonator Configurations}
    %Add-drop resonators (Fig. 2)
    There are two common configurations for passive resonator weights: all-pass and add-drop, consistent with the terminology used in optical filters. In the all-pass configuration, there is only one waveguide bus coupled to the cavity, with one input port and one `thru' port at the opposite end. On the other hand, in the add-drop configuration, an additional bus waveguide is coupled to the cavity, with one `add' port and `drop' opposite to add. In each configuration, the ports are defined relative to the input port. The `thru' port receives most of the energy from the input port, except when the cavity is near resonance. Near resonance, a significant amount of the input energy enters the cavity and gets transferred to the `drop' port. For passive, lossless cavities, all energy from the input port is divided between the `thru' and `drop' ports, in a ratio dependent on the input's wavelength and the coupling between the bus waveguides and the cavity. In theory, the goal is to engineer the resonator so that when on resonance, 100\% of the energy drops to the drop port, and when off-resonance, 100\% of the energy is transmitted to the `thru' port. In a lossless resonator, this can be simply achieved by designing identical couplers between the all-pass and add-drop waveguides, a condition called \emph{critical coupling}.

    In practice, cavities are not perfectly lossless. The main loss mechanisms in silicon photonics are scattering loss (due to roughness of the waveguide's walls) and material absorption. This loss will perturb the critical coupling condition, causing the power transfer to be unbalanced~\cite[Sec. 2]{bogaerts_silicon_2012}. This unbalance can be compensated by design if, and only if, the cavity loss and the coupling coefficients are well known prior to fabrication. These parameters are defined by fabrication and cannot be changed. In a completely passive all-pass microring resonator, for example, this loss is so low that engineering the critical coupling condition is not possible due to fabrication variation.

    We can use a simple technique to achieve near-critical coupling by using a symmetric add-drop bus waveguide, even if we ignore its ports. Because the two couplers are spatially close to one another, the fabrication variation tend to be correlated, ensuring their coupling are similar. This technique only work if the coupling coefficient is much higher than the cavity loss.
    However, as we show in Sec.~\ref{sec:design for manufacturability}, sometimes we intentionally modify the cavity's waveguide via doping or evanescent coupling, increasing the cavity loss to a regime that warrants compensation.
    
\begin{figure}[!t]
    \centering
    \includegraphics[width=\linewidth]{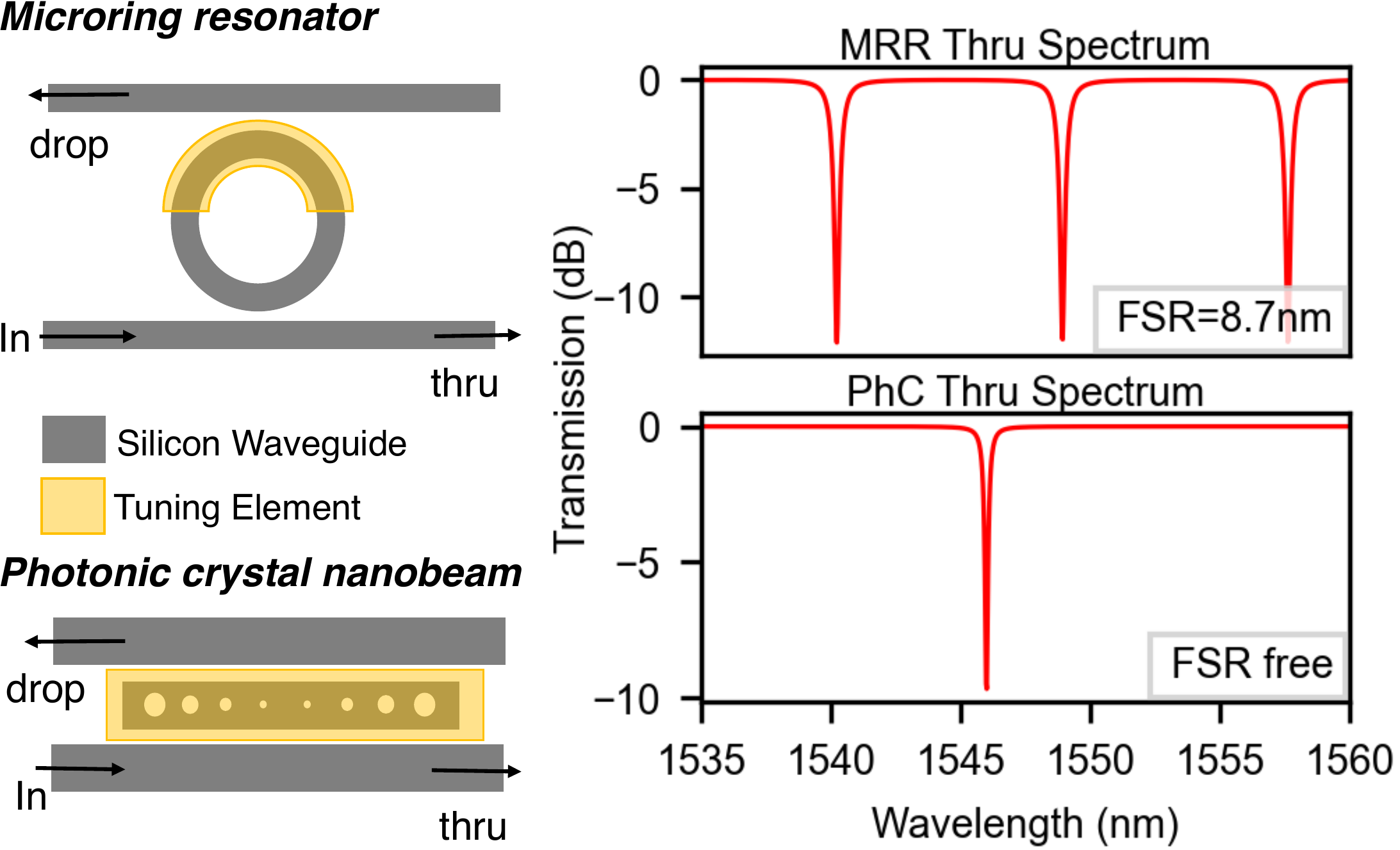}
    \caption{
    \emph{Left:} Top-down view of the microring resonator (top), and the photonic crystal nanobeam cavity (bottom) showing the optical ports and the overlaid area where a tuning element can be placed for controlling resonance wavelength.
    \emph{Right:} Simulated optical spectra as measured from the thru ports in an add-drop MRR (top) and an add-drop PhC nanobeam (bottom) fabricated on a silicon photonic platform. Note that the PhC cavity has a similar resonance feature but no free spectral range (FSR). The parameters used for the MRR were taken from Table~\ref{tab:examples}, and for the PhC from Ref.\cite{zhou_compact_2017}.
    }
    \vspace{-12pt}
    \label{fig:resonators}
    \end{figure}

    The most common example of a photonic resonator is a microring resonator (refer Fig.\ref{fig:resonators}). MRRs are standard components in integrated photonics used for filtering, modulation given their simplistic design methodology. They are ring waveguides where the optical path length, i.e. ring circumference, determines the resonant wavelength. By varying the radius, a WDM link with MRRs with various resonances can be designed. Besides the resonance wavelength, the ring radius also affects the free-spectral range (FSR) of the cavity, which is the spacing between adjacent resonator modes. FSR ultimately is the limiting factor to the channel count of the WDM link. Sophisticated measures have been employed to avoid this FSR limit, including grating-embedded MRR \cite{eid2016fsr}, etc. However, these schemes add to the design complexity and insertion loss. Another way to circumvent this issue would be using photonic crystal (PhC) cavities. Given the relatively strict resonance constraints of PhCs, resonance modes are fewer and can even be engineered to have a single resonance \cite{zhou_compact_2017}. Eliminating FSR offers advantages in terms of channel count, which may be key for large-scale networks. Nevertheless, the fabrication tolerances of PhC's require more expensive fabrication techniques such as E-beam lithography which poses a question to its scalable manufacturability.

    This article was written with microring resonator cavities in mind, but all the discussions and modeling can be applied to any kind of cavity. We have designed a tool to help engineer MRRs based on known fabrication parameters and waveguide designs\footnote{\url{https://github.com/lightwave-lab/microring-resonator-weight-design}} which is also detailed in Sec. \ref{sec:designTool}. Table~\ref{tab:examples} presents example microring resonator designs useful for weighting, and compares their predicted properties to measured properties. With no fit to experiment, the ring properties can be estimated.
    
    \begin{table*}[]
      \centering
      \begin{adjustbox}{max width=\textwidth}
      \begin{tabular}{c|cccc|ccc|c}
      Device & \multicolumn{4}{c|}{Geometry}                & \multicolumn{3}{c|}{Simulated (Measured) parameters} &         \\ \hline
      Type &  \vtop{\hbox{\strut Width}\hbox{\strut (nm)}} & \vtop{\hbox{\strut Radius}\hbox{\strut ($\mu$m)}} & \vtop{\hbox{\strut Gap}\hbox{\strut (nm)}} & \vtop{\hbox{\strut Tuning}\hbox{\strut element}} & \vtop{\hbox{\strut Thru E.R.}\hbox{\strut (dB)}}    & \vtop{\hbox{\strut FWHM}\hbox{\strut (nm)}}   & \vtop{\hbox{\strut Tuning}\hbox{\strut efficiency}}    & Good for            \\ \hline
      Weight     &   500              &     8              &   200    &   n-doped heater    &       22 (25)         &   0.4 (0.4)      &  (0.2) nm/mW   &  Weighting   \\ \hline
      Modulator     &      500           &       8             &  300     &    pn junction   &      13 (15)         &    0.1 (0.03)    &  (0.01) nm/V   &   High-speed tuning    \\
      \end{tabular}
      \end{adjustbox}
      \caption{Example symmetric add-drop microring resonator designs and their uses. Waveguide widths are single-valued across bus and ring waveguide. Doping levels are $\around \SI{e17}{\cm^{-3}}$ and overlapped with the central portion of the rings away from the couplers, leading to a propagation loss of roughly $\SI{5}{\dB/\cm}$~\cite{nedeljkovic_free-carrier_2011}. Parameters were simulated as described in Appendix \ref{sec:designTool} without experimental calibration. We compare them to parameters extracted from measurements taken from fabricated rings.}
      \label{tab:examples}
    \end{table*}
    
    \subsection{Context}
    % Signal vs weight
    \underline{Signals and Weights:} In an analog matrix-vector multiplication, there are two kinds of vectors in a weight bank: the fast vector ($\vec{x}$), which typically represents a signal; and the slow vector, or weight vector ($\vec{w}$), which is understood to be much slower than or stationary compared to $\vec{x}$.
    In digital hardware, this distinction might seem strange because both vectors are typically loaded from memory at the same rate. In analog hardware, however, it is much more energy efficient to use a slow signal to modulate the amplitude of another signal in a passive way. In photonics, for example, a voltage or current-controlled device can modulate the amplitude or phase of an optical signal, independent of its bandwidth. This is the first caveat for comparing performance between photonic matrix-vector multipliers and electronic ones, especially in terms of energy consumption or speed; in analog computing only one operand can be switched quickly, whereas in digital electronics both weight or signal are functionally equivalent.

    One dimension in which optical weights offer a unique advantage is in the latency of the computation. Many applications, such as machine learning, signal processing, and control, require very high bandwidth input signals and low latency but can tolerate slowly-varying weights. For these applications, implementing the MVM in the optical domain is ideal for two main reasons. First, passive optical weights support high bandwidth optical signals with constant (sometimes zero) power dissipation. Second, the latency between input and output is determined by the time-of-flight of the lightwave carrier, or nanoseconds considering resonator and optoelectronic RC-limited bandwidths~\cite{tait_application_2017}.

    Applications such as photonic accelerators, where the computation is offloaded from the CPU to a photonic tensor processor, would not benefit from this scheme since weights would need to be updated as fast as the incoming signals. Doing so imposes a significant energy and speed cost to the control hardware, and is challenging to do in a way that maintains the high-bandwidth advantage of photonics. However, recent works have been able to marry the highly parallel efficient linearity for non-resonant weight elements, and custom digital electronic ASIC for accelerating deep neural networks~\cite{demirkiran_electro-photonic_2021}. To date, this approach has been favored by industry-oriented start-ups.

    % Summing element 
    \underline{Analog Summation:} At the output of the weight bank there is a summing element, most commonly implemented with a photodetector or pair of balanced photodetectors~(cf.~Fig.~\ref{fig:MVM}). Photodetectors generate a photocurrent proportional to the incident optical power. To first order, they are linear devices and a WDM system acts as a multilinear map, meaning that a linear change in each channel results in a linear change in the output. Realistic photodetectors, however, have nonlinearities at high optical powers, resulting in a saturation curve. In practice, this limits the optical power that can be used in a weight bank. They also typically have non-flat frequency response, so optical signals modulated at different frequencies but with same power generate different levels of photocurrent. Any nonlinear imperfection in the photodetector will contribute to precision and accuracy `errors' in the weighting scheme.

    \underline{Analog Subtraction:} It is often desirable to implement negative weights (in addition to positive weights) in order to allow for the possibility of analog subtraction. Doing so in a WDM system requires the use of add-drop resonators along with balanced detectors, as depicted in Fig.~\ref{fig:MVM}. In such a case, a channel's weight value is zero when energy is evenly split between ``thru'' and ``drop'' ports.
    
    \underline{Mach-Zehnder Interferometer (MZI) Approaches:} An alternative to WDM-based integrated photonic computational techniques are techniques which utilize Mach-Zehnder Interferometers (MZIs) to perform arbitrary linear operations on a single-wavelength optical carrier \cite{harris_linear_2018}. Individual phase shifters found in MZIs are typically less sensitive than individual resonators, but process variations compound across large many-MZI PICs in a way that is not usually seen in large many-resonator PICs. Ultimately, compensating calibration and control techniques are required for practical deployment in either case. A discussion of the most common options for MZI-Based PICs can be found in \cite{harris_linear_2018}.
    
    \subsection{Definitions}
      {\underline{Normalized Weights}}: For simplicity's sake, we introduce a normalized weight $\hat{w}\in[0,1]$. Conversion between real weight $w$ (as measured by the analog summing element) and normalized weight can be accomplished via the equation:
      
      \begin{equation}
          \hat{w}=\frac{w-w_{\text{min}}}{w_{\text{max}}-w_{\text{min}}}
      \end{equation}
      
      where $w_{\text{max}}$ and $w_{\text{min}}$ are the maximum and minimum weight values respectively.

      {\underline{Bit Representation}}: Before defining error quantities such as accuracy, precision and resolution, sometimes it is useful to refer to them in units of `bits'. For example, if the relative error of a measurement is \num{0.125}, or \SI{12.5}{\percent}, we can also refer to it as \SI{3}{bits}, because it takes three digits to represent that number in binary representation. More generally, if the error is \(\varepsilon \in (0,1]\), the bit representation can be computed as \(\varepsilon \,\textrm{(bits)} = \log_2(1/\varepsilon) \in [0, \infty)\). For example, as the measurement error goes to \num{0}, we say that it has infinite precision.
    
      {\underline{Accuracy}}: In the context of resonator weight banks, accuracy refers to the systematic error between the commanded weight (denoted \(\hat{\vec{w}}\)) and actual resulting weight vector (\(\vec{W}(\hat{\vec{w}})\), where \(\vec{W}\) is a random variable for commanded weight \(\hat{\vec{w}}\)). Here we make the distinction between individual accuracy, which refers to accuracy of any given individual weight, and ensemble accuracy, the average accuracy over all possible individual weights. From a user's perspective ensemble accuracy is a more useful metric, since it reflects expected performance in the general case where weights are unknown ahead of time. Accordingly, we refer to ensemble accuracy wherever it is not explicitly stated.
      In theory, if laboratory conditions stay stationary over time and instruments are perfectly repeatable, a perfect calibration algorithm would yield infinite ensemble accuracy (zero error). Such an algorithm could visit every possible commanded weight combination, measure the effective weight, and construct a lookup table with the corresponding map. In practice, however, it is desirable to perform this calibration procedure as quickly as possible and with a model that is simple enough to implement in a microcontroller. Ensemble accuracy, therefore, is a metric that measures the quality of the instrumentation, the correctness of the physical model and the efficiency of the calibration algorithm. Accuracy is defined mathematically as:
      
      \begin{align}
        \text{Individual Accuracy } \delta_{\hat{\vec{w}}} &\triangleq \norm{\hat{\vec{w}} - \mathop{\mathbb{E}}[\vec{W}(\hat{\vec{w}})]}\\
        \text{Ensemble Accuracy} &\triangleq \sqrt{\sum_{\hat{\vec{w}} \in \vec{\Omega}}\delta^2_{\hat{\vec{w}}}},
      \end{align}
      where \(\norm{\cdot}\) denotes the euclidean norm, \(\mathop{\mathbb{E}}[\cdot]\) denotes the expected value, and \(\sum_{\hat{\vec{w}} \in \vec{\Omega}}\) denotes the average across all possible weight vectors.

      {\underline{Precision}}: Precision is complementary to accuracy, and refers to random error caused by noise in the measurement system or control circuit. This is a challenge in any analog system. Here we again make the distinction between individual precision and ensemble precision, and in general refer to ensemble precision whenever it is not specified. In this case, we chose to construct the ensemble precision so that it includes a contribution from individual accuracies in order to more closely reflect the user's experience. This is intuitively understandable by examining each limiting case. On one extreme, if all individual accuracies are infinite (zero error) then random spread of the ensemble is just the average of individual precisions. On the other extreme, if all individual precisions are infinite (no random spread) then one still expects to see spread in commanded weight values over the entire ensemble due to finite accuracy. Here, we associate ensemble precision to an error bound users would expect from a randomly chosen weight vector. Mathematically, individual and ensemble precision are defined as:
      
      \begin{align}
        \text{Individual Precision } \sigma_{\hat{\vec{w}}} &\triangleq \sqrt{\Var{\vec{W}(\hat{\vec{w}})}}\\
        \text{Ensemble Precision}&\triangleq \sqrt{\sum_{\hat{\vec{w}} \in \vec{\Omega}}\left(\sigma_{\hat{\vec{w}}}^2 + \delta_{\hat{\vec{w}}}^2\right)},
      \end{align}
      with the same notation as in the accuracy definition.

      {\underline{Resolution}} ordinarily relates to the discretization error of the signals \(x_i(t)\) or \(y(t)\), which is not to be confused with weight accuracy and precision. For analog continuous signals, the equivalent terminology is signal-to-noise ratio, or SNR. In this article, we use resolution and SNR interchangeably, often expressed in bits. Because of Eq.~\ref{eq:vectordotproduct}, the finite precision error of \(\vec{w}\) causes a degradation in the resolution of \(y(t)\). In practice, weight banks are designed such that the resolution is lower than the ensemble precision (e.g. 5-bit precision for a 4-bit resolution), otherwise ensemble precision will determine the resolution of \(y(t)\).

      {\underline{Actuation}} refers to the physical process of changing a weight. One common actuation mechanism is the application of current to a microheater near a resonator. Applied current generates heat, which changes the temperature around the resonator and as a result changes the weight. The actuation mechanism often sets an upper limit on achievable accuracy. For example, consider a resonator controlled by a typical 16-bit DAC within a \SIrange{0}{20}{mA} range. If the resonator needs \SI{2}{mA} of current to lock to resonance and \SI{1.5}{mA} to change the effective weight between -1 and 1, a single LSB of the DAC corresponds to \SI{0.3}{\uA} and $-\log_2(\SI{0.3}{\uA}/\SI{1.5}{mA})=\SI{12.3}{bits}$ is the maximum achievable accuracy. In practice, DAC nonlinearities, parasitic resistances, and on-chip thermal crosstalk tend to decrease system accuracy below this upper limit.

      {\underline{Sensing}} is the ability to directly measure each effective weight independently from the actuation mechanism. This plays an important role in simplifying the calibration and control of the weight bank as well as improving accuracy and reconfiguration speed. 
      The ideal sensor would be a spectral sensor that would detect the transmission of each resonator as a function of wavelength and map it to a voltage with high accuracy (0 to 10V would be compatible with popular DAC/ADC components currently available from chip manufacturers).
      More practical sensors measure indirect quantities correlated with the weight, such as local temperature or circulating optical intensity. In this case, the relationship must be incorporated into the calibration and control.

    %   \fxwarning{The following could probably be moved}. With a series of dedicated sensors, a complete physical model of each resonator and careful monitoring of environmental conditions are no longer necessary for reliable operation of the weight bank. As we will show later, a calibration model and some monitoring are still necessary if the sensor indirectly detects the applied weight. This situation is common in practical systems, as it is much simpler to implement an accurate sensor for a correlated physical quantity, such as local temperature or optical intensity, rather than optical spectrum or transmission itself.

      {\underline{Trimming}}: refers to the process of adjusting a physical property of a device to a desired value with high accuracy. This is necessary because common microfabrication processes based on photolithography or e-beam lithography are fundamentally incapable of producing predictable, identical resonator cavities (thus resonance wavelengths) by design. For example, a typical requirement in EDM telecomm. systems is a fixed operating wavelength, say \SI{1545.32}{\nm} (ITU C-band channel 40). A resonator designed to operate on this channel needs a resonance wavelength accuracy  of \SI{0.01}{\nm}, which is currently unattainable by geometric design alone. 
      There are two main ways of achieving trimming post-fabrication. Active trimming involves measuring each resonator's wavelength in advance and using a control circuit to apply a voltage or current bias that compensates for fabrication variations. This approach is referred to as active because it requires the use of external powered circuitry to apply the correction. In contrast, passive trimming involves permanently adjusting the device to the target wavelength via modifications to its physical properties (e.g., phase-change materials or ion implantation). Passive approaches are preferred because they reduce extra control logic and power requirements. Passive trimming technologies, however, currently require specific post-processing steps that may not be economically viable in silicon photonic foundries.
      
      It is worth noting that active trimming and weighting occur via the same physical mechanisms. The difference between the two therefore exists solely in the software layer (i.e., how the user interacts with each). An end user should only need to control weighting, and should not need to worry about active trimming. In other words, the same weight command should produce the same results in two separate chips regardless of manufacturing imperfections. Trimming, therefore, must be completely automatic and free from user interaction. This separation is absolutely critical to enable use of photonic computing hardware by non-experts, and consequently for the proliferation of the technology. In order to realize this, there should be two parallel circuits that are independent and are characterized by independent metrics: trimming, which makes sure that any chip with the same design has the same baseline behavior; and weight control, which translates weight command to effective weight.

\section{Control Algorithms}
    \label{sec:calibration and control}

    One of the biggest advantages of using analog photonic weights is that high-bandwidth optical signals can be weighted by an optoelectronic device that consumes constant power (or static power).
    % Tait's comment: This is a bit strong. You need enough light to do weighting. Higher bandwidth means more noise, meaning more signal, more light. Lower bandwidth means that number of operations drops relative to static power.
    It is worth noting however that as bandwidth is increased, more static power is required at the optical source to compensate for the added noise captured by the summing photodiodes.
    In the mechanisms presented in the previous section, the analog weight is defined by the electrical voltage or current applied to the microresonator.
    The accuracy and precision of each weight is a function of the completeness of the model along with the effects of noise and disturbances. In this section, we outline a simplified weight bank model and discuss practical examples of noise and disturbances.

    \subsection{Simplified Microresonator Weight Bank Model}
    \label{subsec:simplified model}
    
    To illustrate the ramifications of control algorithm, we present a simplified model of a resonator weight bank (Table~\ref{tab:simplified_model}). Each weight (e.g. a microring resonator) can be tuned via silicon's thermo-optic effect by controlling the local temperature. By sourcing current through a resistor placed nearby, this temperature can be adjusted directly and efficiently. This phenomenon is referred to as Joule heating.
    We assume, for simplicity, that the properties of the resistor do not change with its temperature or with the circulating optical power in the resonator cavity.
    %Joule heating will cause a temperature gradient around the resistor, increasing the average temperature of the microresonator. \fxwarning{The temperature shift, thermo-optic component, and Joule heating are all mentioned twice. This should be consolidated}. Then, in silicon resonators, this temperature shift will cause a change in refractive index of the waveguide. This is called the thermooptic effect, and is very strong in silicon.
    Due to the phase-matching condition on resonators, there is a direct relationship between resonant wavelength and refractive index, namely \(\lambda/n = \textrm{constant}\).
    In a simple model, the transmission of the microresonator is a function of only the ``detuning,'' or the difference between the fixed wavelength (\(\lambda_0\)) and the tunable resonance frequency (\(\lambda\)). In high Q-factor resonators, this function can be approximated by a Lorentzian line shape. In Table~\ref{tab:simplified_model}, this is denoted by \(\mathcal{L}(\lambda - \lambda_0)\).
    The second-order deviations from this model are: dependence of transmission on optical intensity (at \(\lambda_0\)), optical crosstalk from neighboring resonators, and undesired optical attenuation in the presence of electrical carriers (most important for doped waveguides). 
    
    After weighting, the output of the weight bank is fed to a photodetector. In our simplified model, this photodetector is assumed to have a constant responsivity independent of input power. The electrical properties of the photodetector (parasitic capacitance, parasitic resistance, or frequency response) can be ignored at this point and instead captured later by the model of the electrical receiver (in neuromorphic photonics, this would be the E/O conversion of the neuron). In practice, the photodetector's responsivity can have a range of values depending on fabrication variation and operating wavelength, and this should be captured by the calibration model via a linear correction. Once the optical power input is too high, however, the responsivity drops due to space-charge screening effects~\cite{williams_effects_1994}. This nonlinear relationship between photocurrent and optical intensity cannot be calibrated away, and results in reduced precision.

    % more sophisticated models were developed to yield better accuracy~\cite{Tait2016}.
    
    \begin{table}
    \caption{Simplified Microresonator Weight Bank Model}\label{tab:simplified_model}
    \begin{algorithmic}[1]
    \State \textbf{Joule Heating:} 
      \[T_i = T_{0} + \sum_j^N\mathbf{K}_{ij} R_j I_j^2\]
      
      \Comment{ $T_i$: Local temperature; $T_0$: Room temperature; $\mathbf{K}$: Effective thermo-optic coefficient matrix (diagonal in the absence of thermal crosstalk); $R_i$: Electrical heater resistance.}

    \State 
      \textbf{Thermooptic effect:}
      \[\lambda_i = \lambda_{0,i} \left(1+\frac{\beta_{\mathrm{eff},i}\Delta T_i}{n_\mathrm{eff}(\lambda_{0,i})}\right)\]
      
      \Comment{\(\lambda_i\) (\(\lambda_{0,i}\)): resonance frequency of filter with applied heat (at room temperature), respectively; \(\beta_{\mathrm{eff}}\): thermo-optic coefficient; \(n_\mathrm{eff}(\lambda_{0,i})\): effective waveguide index; \(\Delta T\): temperature offset.}

    \State 
      \textbf{Lorentzian transfer function:}
      \[\mathcal{L}(\lambda; \lambda_0, \Delta\lambda) = 1 - \left(1+(\lambda - \lambda_0)^2/\Delta\lambda^2\right)^{-1}\]

      \Comment{\(\Delta\lambda\): Half-width half-maximum of Lorentzian transfer function modeling a single resonator weight.}

    \State 
      \textbf{Weighted sum}
      \[\mu = \frac{1}{N}\sum_i^N \mathcal{L}(\lambda_{l,i}; \lambda_{0,i}, \Delta\lambda_i) \cdot P_i/P_\text{max} \]

      \Comment{\(\mu \in [0,1]\): weighted sum (photocurrent normalized by PD's responsivity); \(P_i\): optical intensity of input lightwave at wavelength \(\lambda_{l,i}\); \(P_\text{max}\): maximum optical intensity.}
    \end{algorithmic}

    % \begin{tabularx}{\linewidth}{|c|X|c|}
    % Symbol & Description & Unit \\
    % $T_0$ & Room temperature & \unit{\degreeCelsius} \\
    % $\mathbf{K}$ & Effective thermo-optic coefficient matrix. Mostly diagonal. &\\
    % $\lambda_0$ & Resonance frequency of filter at $\Delta T=0$. & \unit{nm}\\
    % $\lambda$ & Resonance frequency of filter with applied heat. & \unit{nm}\\
    % $\lambda_l$ & WDM laser frequency & \unit{nm} \\
    % $\Delta\lambda$ & Half-width half-maximum of filter. & \unit{nm} \\
    % $\beta$ & Thermo-optic coefficient & \unit{\degreeCelsius^{-1}} \\
    % \end{tabularx}
    \end{table}

    \subsection{Compensating for Noise and Disturbances}
    \label{subsec:impacts of noise}

    The simplified model presented in Sec.~\ref{subsec:simplified model} is effective at modelling the weighted sum \(\mu\) as a function of the optical intensities at each wavelength \(P_i\). In this scheme, the signal is modulated onto each carrier wavelength. To simplify, we assume a perfect modulation scheme where the instantaneous intensity is described as follows:

    \begin{equation}
    P_i(t) = P_{i,\textrm{input}}x_i(t),
    \label{eq:intensitymodulation}
    \end{equation}
    where \(x_i(t)\) is the same as in Eq.~\ref{eq:vectordotproduct} and \(P_{i,\textrm{input}}\).

    In this analog representation, we can identify \(\mu(t) / P_0\) as the equivalent of the weighted sum \(y(t)\) scalar, and the weight as: 
    \begin{equation}
    w_i \equiv \mathcal{L}(\lambda_{l,i}; \lambda_{0,i}, \Delta\lambda_i) \cdot P_{i,\textrm{input}} / N P_0,
    \label{eq:weight analog}
    \end{equation}
    where \(P_0\) is a normalization constant.

    As revealed by Eq.~\ref{eq:weight analog}, however, noise and temporal disturbances to the values of \(P_{i,\textrm{input}}\) can negatively affect the weight's precision or the resolution of the sum \(y\). These effects cannot be captured by any calibration model, since by definition a calibration model must be based on static parameters. Disturbances must nonetheless be corrected independently from the transmission control, since the user will reasonably expect the chip to have a constant average weight vector.

    \begin{figure}[ht]
      \centering
      % \includegraphics[width=\linewidth]{box2-photonic-link}\\
      % Sources of fluctuation in the signal pathway of a typical O/E photonic link.
      % \[
      % \underbrace{\frac{\delta I_\mathrm{out}(t)}{I_\mathrm{out}}}_{\text{real signal}}
      % =
      % \underbrace{\frac{\delta_t P_\mathrm{pump}}{P_\mathrm{pump}}}_{\substack{\text{light source} \\ \text{fluctuation}}}
      % -\underbrace{\delta_t \alpha_\mathrm{GC}}_{\substack{\text{coupling} \\ \text{fluctuation}}}
      % +\underbrace{\frac{\delta_{I,\theta} T(I_j, \theta_\mathrm{amb})}{T}}_{\substack{\text{linear filter} \\ \text{fluctuation}}}
      % +\underbrace{\frac{\delta_t x(t)}{x}}_{\text{true signal}}
      % \]
      \includegraphics[width=\linewidth]{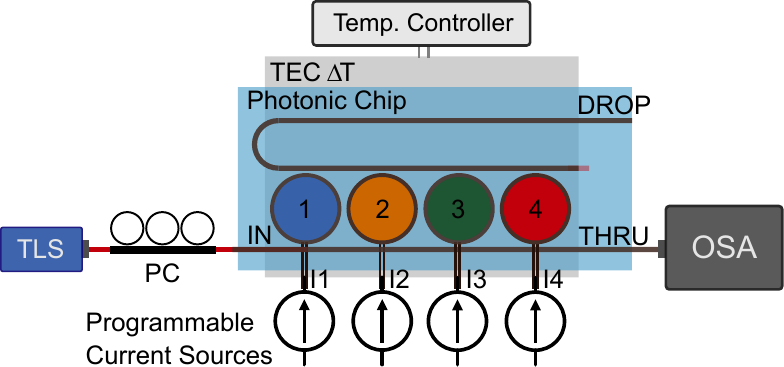}\\
      \caption{
      Experimental setup of an optical transmission spectrum measurement 4-neuron microring resonator weight bank. The PIC sits atop a thermoelectric module whose temperature can be electrically controlled. A tunable laser source (TLS) generates continuous-wave laser light whose wavelength is quickly scanned. An optical spectrum analyzer (OSA) synchronized with the TLS collects snapshots of the transmission spectrum. A polarization controller (PC) is used to emulate large polarization drifts due to long-term environmental fluctuations.
      }
      \label{fig:MWB_setup}
    \end{figure}

    \begin{figure}[ht]
      \small
      \centering
      \small
      % \includegraphics[width=\linewidth]{box2-photonic-link}\\
      % Sources of fluctuation in the signal pathway of a typical O/E photonic link.
      % \[
      % \underbrace{\frac{\delta I_\mathrm{out}(t)}{I_\mathrm{out}}}_{\text{real signal}}
      % =
      % \underbrace{\frac{\delta_t P_\mathrm{pump}}{P_\mathrm{pump}}}_{\substack{\text{light source} \\ \text{fluctuation}}}
      % -\underbrace{\delta_t \alpha_\mathrm{GC}}_{\substack{\text{coupling} \\ \text{fluctuation}}}
      % +\underbrace{\frac{\delta_{I,\theta} T(I_j, \theta_\mathrm{amb})}{T}}_{\substack{\text{linear filter} \\ \text{fluctuation}}}
      % +\underbrace{\frac{\delta_t x(t)}{x}}_{\text{true signal}}
      % \]
      \includegraphics[width=\linewidth]{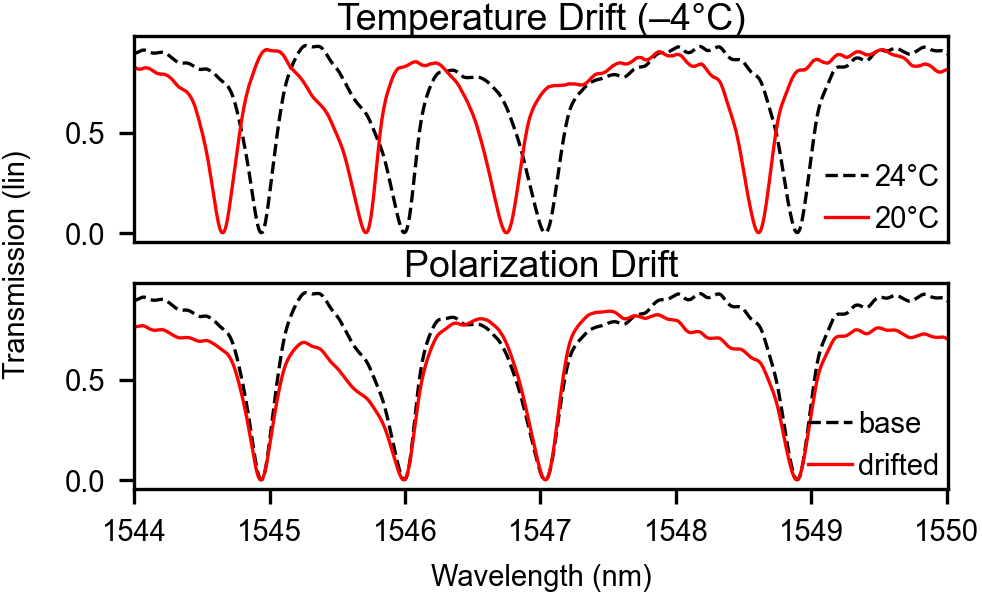}\\
      \caption{
      Experimental results of a weight bank's transmission profile when subject to a temperature drift (top) or an input polarization drift (bottom). When the stage was cooled from \SI{24}{\celsius} to \SI{20}{\celsius}, the resonance wavelengths of each ring blue-shifted, as expected from the positive coefficient of the thermo-optic effect in silicon. The polarization drift was induced by rotating the planar polarization angle between the laser and the input coupler (cf.~\ref{fig:MWB_setup}), resulting in imperfect coupling to the chip and unpredictable changes to the transmission spectrum.}
      \label{fig:polarization and temperature drift}
    \end{figure}

    In practice, there are two relevant sources of disturbances to the input intensity. One is ambient temperature, and the other is polarization drift accumulated in the input fibers. To show the impact these sources have on the operation of a weight bank, we have setup a microring resonator weight bank with four rings, placed on top of a temperature controlled base (Fig.~\ref{fig:MWB_setup}). These effects can be observed, for instance, in the optical spectrum of a microring resonator weight bank (Fig.~\ref{fig:polarization and temperature drift}). We observe that the temperature drift essentially maintains the shape of the transmission curve for each resonator, as expected for microring resonators. As a result, the microcontroller can compensate for that by either dissipating extra heat uniformly across all devices, or by using a thermal heat bath that holds the average temperature of the entire chip constant. This invariance property is unique to microrings and cannot be generalized to other kinds of resonator structures.
    A polarization drift in the fibers outside the chip, however, alters the power spectrum irregularly, and can only be remedied by recalibrating the circuit. We have found that using polarization-maintaining fibers and optical adhesives can stabilize the polarization spectrum indefinitely in laboratory settings. More studies must be completed in order to evaluate this effect in field conditions, where significant contributions are expected from mechanical stress and vibrations.
    % \fxnote{SB: how much of the polarization issue is from the resonator vs the rest of the circuit e.g. grating couplers? TFL: Completely outside of the purview of the weight bank. GCs, fibers, discrete components' fault...}

    % \fxwarning{Not all problems can be controlled. There are many sources of fluctuation that cannot be mitigated by smart actuation and sensing schemes. See Fig.~\ref{fig:polarization and temperature drift}. These terms should be understood and reported as noise.}

    % Fig.~\ref{fig:polarization and temperature drift} shows experimental optical spectra of a microring weight bank of 4 rings subject to a temperature drift and a polarization drift. \fxwarning{describe experiment to obtain those. Remote top figure but describe in words here.}

   % Measuring weights
    \subsection{Measuring Effective Weights} 
    The summing element is an integral part of the photonic circuit, so we typically only have access to output photocurrent while operating the weight bank. This creates the challenge of independently verifying that the applied weight was, in fact, the correct one. Mathematically speaking, in the linear approximation of the summing element, for a set of signals \(x_i(t)\), and a measurement of \(y(t)\), we need to recover the weights \(w_i\) in Eq.~\ref{eq:vectordotproduct}. One approach could be to ensure that only one channel has a positive value at any given time, for example \(x_i(t) = 1\) for time slot \(t \in \left(i\Delta t, (i+1)\Delta t\right]\) and \(x_i(t)=0\) otherwise. Then, the measurement of \(y(t)\) yields \(y(i\Delta t) = w_i\). A better approach, which is more often used, is to use independent (orthogonal) signals e.g. \(x_i(t) = \sqrt{2}\sin(\omega_i t)\) which have the property that \(\int x_i(t)x_j(t) = \delta_{ij}\). Then, a single measurement of \(y(t)\) can be used to compute all weights with the following decomposition: \(w_i = \int y(t) x_i(t)\). With this method, \(n\) weights can be continuously monitored for every integration step.
    
    \subsection{Calibration and Feedforward Control}
    \label{subsec:feedforward control}

    Regardless of the weighting mechanism, the control strategy of a resonator involve a calibration stage and a control stage. The calibration is required to adjust the weight bank model parameters caused by fabrication variation or uncertain initial conditions. The control stage runs continuously thereafter.

    \begin{figure}[ht]
    \centering
    \includegraphics[width=\linewidth]{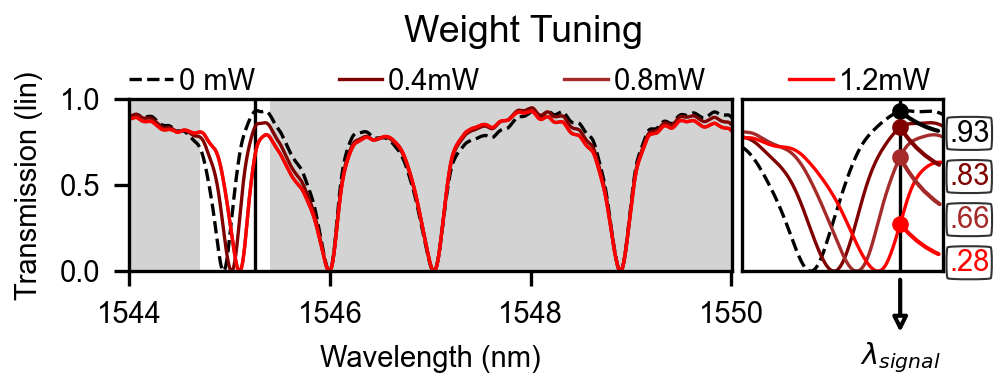}\\
    \caption{Experimental optical spectral response of a four-ring weight bank as current flows through the heater situated on top of the first ring in the bank (cf.~\ref{fig:MWB_setup}). To first order, only one resonance feature is shifted, without affecting others. The inset on the right corresponds to the unshaded area of the left, and shows the transmission values for each of the four tuning powers shown in the legend above. Note that the resonance shift is approximately proportional to the electrical power dissipated on the heater element.
    }
    \label{fig:resonator-weighting}
    \end{figure}

    Calibration and control strategies must take into consideration an array of resonators in the weight bank, because coupled resonators suffer from both optical and electrical crosstalk. As shown in Fig.~\ref{fig:resonator-weighting}, applying Joule heating to the vicinity of one resonator directly changes its transmission function, which according to the model introduced in Sec.~\ref{subsec:simplified model} results in changing the weight for that wavelength. Notice, however, that the neighboring resonator also undergoes a small change in its transmission function. This is known as crosstalk. Though barely visible, this second order effect, if uncompensated, jeopardizes the accuracy of the weight bank. For a 8-bit accuracy level, the transmission on the other rings must be kept within 0.4\% of their target. In our model, we can capture this crosstalk in the matrix $\mathbf{K}$, shown in Table~\ref{tab:simplified_model}.

    As the number of weights \(N\) in a system increase, so do the number of independent variables in the model. Mitigating the electrical and optical crosstalk between resonators in a weight bank requires measuring crosstalk terms and fitting a multidimensional model with \(\mathcal{O}(N^2)\) variables. For example, the crosstalk matrix $\mathbf{K}$ is of dimension $N\times N$.
    % To make things worse, the resonators' add-drop ports are optically connected with each other, reducing the number of optical ports available for calibration.
    Consequently, we must rely on indirect measurements and physical models to control each resonator in the weight bank system.

\begin{figure}[!ht]
      \centering
      \includegraphics[width=\linewidth]{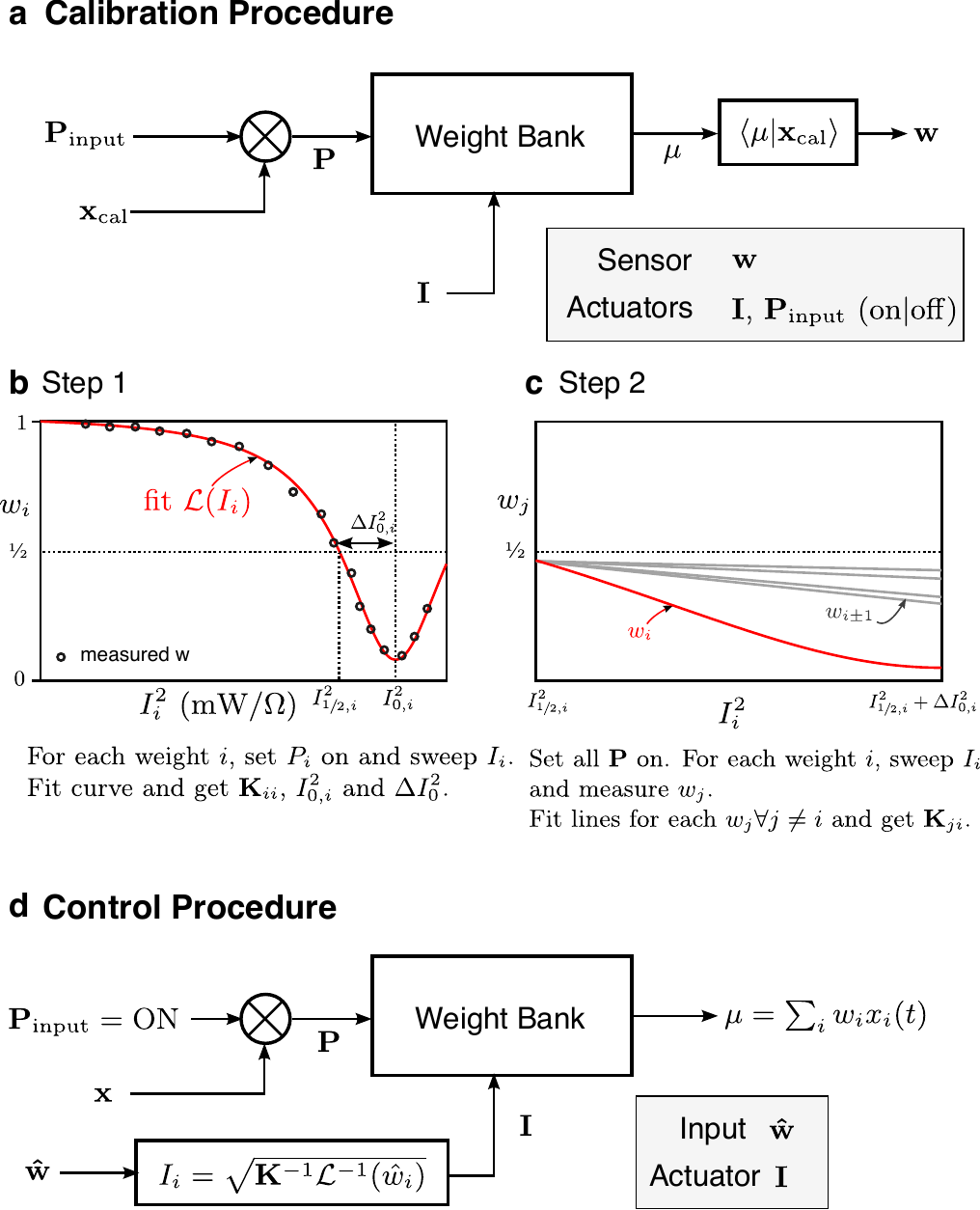}\\
      \caption{Procedure for feedforward calibration and control. The calibration can be performed in a single sweep for each resonator (\(\mathcal{O}(N)\)) with a specially crafted calibration signal $x_\text{cal}$ (a). The calibration must be completed in two steps (b,c). Each step completes in \(\mathcal{O}(N)\) substeps, if we count that the \(\mathbf{w}\) vector is measured in a single substep (c). Once the calibration is completed and the crosstalk matrix \(\mathbf{K}\) is known, the control step uses the stored model to calculate the desired electrical current inputs (d). The overall computational complexity is \(\mathcal{O}(N)\).
      This scheme requires another calibration procedure anytime the chip undergoes any environmental variation such as a temperature shift.
      }
      \label{fig:feedforward calibration and control}
    \end{figure}

    A series of approaches have been developed to tackle the weight bank control problem. The earliest method presented in the literature is called the feedforward scheme~\cite{tait_multi-channel_2016}, shown in Fig.~\ref{fig:feedforward calibration and control}. In this scheme, the controller would possess an electrical model of the resonators, including an approximation of the thermal and optical crosstalk effect. The model maps a vector of electrical current values to a vector of weight values, and vice versa. Based on a desired weight, the model converts to an actuation electrical signal for each resonator. But once signals are set, there is no way to verify or validate that the effective weight is correct.

    % \begin{table}[h!]
    % \begin{algorithmic}[1]
    % \Procedure{Calibration}{$\mu,I_i,P_i$}
    %   \State Sensors: \(\mu\) \Comment{Weighted Sum.}
    %   \State Actuator 1: \(I_i\) \Comment{Weight $i$'s heater.}
    %   \State Actuator 2: \(P_i\) \Comment{Laser power at wavelength \(\lambda_{l,i}\).}
    %   \For{$i\gets 1, n$}
    %     \State Set \(P_i = 1\), all others to \(0\).
    %     \State Set \(I_i = 0 ~\forall i\).
    %     \State Sweep \(I_i\) from \(I^0\) to \(I^\textrm{max}\).
    %     \State Measure \(\mu(I_i)\)
    %     \State Fit \(\mathcal{L}(I_i)\) $\rightarrow$ Obtain \(\widetilde{\lambda}_{0,i}\) and \(\widetilde{\Delta\lambda_i}\)
    %       \Comment{First-order estimates of \(\lambda_{0,i}\) and \(\Delta\lambda_i\)}
    %     Set \(I_i = \frac{\Delta I}{K_{ii}}\sqrt{\frac{1-\widetilde{\lambda_{l,i}}}{2}}\)
    %     \For{$j\gets 1, n$}
    %       \State Sweep \(I_j\) from \(I^0\) to \(I^\textrm{max}\).
    %       \State Measure \(\mu(I_j)\)
    %     \EndFor
    %   \EndFor
    % \EndProcedure

    % \end{algorithmic}
    % \end{table}

    The advantage of this approach is the simplicity of the control algorithm -- it is similar to a lookup table in electronics. This calibration algorithm scales with \(\mathcal{O}(N)\), despite the fact that the feedforward model has $\mathcal{O}(N^2)$ parameters. However, this approach has two main disadvantages: the calibration step is complex and does not work when the outputs of the weight bank is not accessible. It also fails to correct for environmental variations or laboratory conditions. As a result, the chip needs to be recalibrated before each use, making this approach impractical for large $N$.

    To resolve these fundamental issues with the feedforward scheme, the microresonator device can be designed with an embedded sensor, capable of measuring the applied weight in real time. With a sensor located near each resonator in the weight bank, measuring e.g. the local temperature, we can directly feed an electric signal back to the controller to adjust for model deviations. This is called a feedback scheme, and will be revisited in Sec.~\ref{subsec:feedback control}, after we delve into index modulation, actuation and sensing physics in photonic waveguides in Sec.~\ref{sec:design for manufacturability}.

\section{Design for Manufacturability \label{design}}
    \label{sec:design for manufacturability}
    % Fabrication imperfections
    Passive optical devices are sensitive to the effective refractive index of waveguides, which determines the optical path length along the path of propagation. In integrated photonics platforms, the effective refractive index varies with the height and width of the waveguide \cite{selvaraja_subnanometer_2010}. The width of integrated waveguides is lower-bounded by a photolithography process with a limited resolution. In the case of silicon photonic chips, deep ultraviolet photolithography is used, with a wavelength of \SI{193}{nm}, limiting the minimum lateral feature size of devices to \SI{65}{nm}~\cite{rahim_open-access_2018}. Wider waveguides are typically used, however, since increasing width leads to tighter optical mode confinement and lower losses. This, in turn, is upper-bounded by the width at which the waveguide supports a second lateral mode at the wavelength of operation. As a result waveguide width is usually chosen to be the maximum single-mode width minus some engineering margin. For similar reasons, the height of the waveguide is often chosen to be the maximum that only admits a single infinite-slab mode minus some engineering margin. A de facto industry standard height is \SI{220}{nm} with a standard deviation of \SI{2}{nm} \cite{selvaraja_subnanometer_2010}.
    
    % Layout alone is not sufficient to mitigate imperfections. Need trimming.
    The combination of lateral and vertical waveguide manufacturing imperfections results in fixed, random phase shifts in each waveguide both across the same chip and between chips. That means that identically-designed interferometric devices will behave differently once fabricated. For resonator structures, random phase offsets above \(2\pi\) are sufficient to render any prediction of a post-fabrication resonance wavelength impossible. However, within the same chips the phase offsets are spatially correlated \cite{chrostowski_impact_2014}, meaning that resonators in close proximity have resonance offsets similar in sign and magnitude.

    % Trimming
    There are several post-fabrication options available to compensate for undesired random phase shifts. These techniques are referred to as trimming, where a section of an optical waveguide material is physically or chemically altered until the desired phase is achieved. The range of phase shifts that can be compensated is typically limited, and as a result design-level compensation is often required. Design for manufacturability tackles this challenge by predicting the amount of compensation required via Monte Carlo simulations of fabrication variability~\cite{lu_performance_2017}.
    A successful design for manufacturability is able to reduce the range of fabrication variation that will need to be compensated by trimming, and can result in considerable cost saving overall.

    \subsection{Index Tuning Mechanisms}
    % Number of refractive index mechanisms Table I
    Many refractive index tuning mechanisms have been developed for silicon photonics, and Table~\ref{tab:modulation-options} displays some state-of-the-art devices from the literature.
    In silicon, the strongest effects are the thermo-optic effect, free-carrier absorption and free-carrier dispersion (also known as plasma dispersion).
    Exploiting the thermo-optic effect for tuning with metal filament microheaters is the easiest and most popular way to effect large index changes, but it is slow and power inefficient~\cite{Jacques2019}. Thermal tuning with waveguide-embedded heaters is similar in efficiency~\cite{Jayatilleka2015}, but provides a potential feedback signal for weight control~\cite{Tait:18fb}. To exploit free-carrier effects, we can directly manipulate carrier concentrations by selectively p- and n-doping the waveguide in a lateral junction~\cite{Patel2015}. Free-carrier absorption dominates if the junction is forward biased, while free-carrier dispersion dominates if the junction is reverse biased.

    \begin{table}[t]
      \small
      {\centering
      \includegraphics[width=\linewidth]{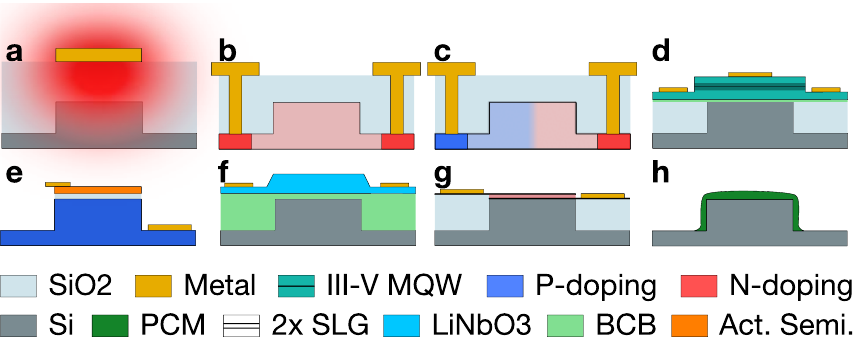}
      \begin{adjustbox}{max width=\linewidth}
      \begin{tabular}{lllc}
        \toprule
        \textbf{Modulation Effect} & \textbf{Speed} & \textbf{Efficiency}         & \textbf{Ref.}              \\
        \midrule
        Thermo-optic TiN (a)    & \around \SI{5.6}{\us} & $P_\pi L = \SI{6.8}{mW.mm}$ & \cite{Jacques2019}               \\
        Thermo-optic N\textsuperscript{+}/N/N\textsuperscript{+} Si (b) & \around\si{\us}         & $P_\pi L = \SI{0.8}{mW.mm}$ & \cite{Jayatilleka2015}               \\
        Reverse-biased PN (c)   & \SI{41}{GHz}         & $V_\pi L = \SI{46}{V.mm}$    & \cite{Patel2015}       \\
        Graphene SLG (g)       & \SI{30}{GHz}$^1$     & $V_\pi L = \SI{28}{V.mm}$    & \cite{Sorianello2018}  \\
        \ch{LiNbO3}/Si Hybrid (f)      & \SI{70}{GHz}         & $V_\pi L = \SI{22}{V.mm}$    & \cite{He2019}  \\
        III-V MQW/Si Hybrid (d) & \SI{27}{GHz}         & $V_\pi L = \SI{2.4}{V.mm}$   & \cite{Chen2011}       \\
        III-V/Si MOS (e)       & \SI{2.2}{GHz}        & $V_\pi L = \SI{0.9}{V.mm}$   & \cite{Hiraki2017}   \\
        ITO MOS (e)            & \around\si{GHz}$^2$  & $V_\pi L = \SI{0.52}{V.mm}$  & \cite{Amin2018}          \\
        Forward-biased PIN (c)  & \SI{0.5}{GHz}        & $V_\pi L = \SI{0.36}{V.mm}$  & \cite{Green2007}       \\
        PCM (h)                & \SI{0.8}{GHz}$^3$    & $E_\pi = \SI{400}{pJ}$  & \cite{Rios2015} \\
        Ion implantation & 0 & 0 & \cite{milosevic_ion_2018} \\
        \bottomrule
      \end{tabular}
      \end{adjustbox}
      }
      %references from 2020-Nature-Photonics-Review
      ~\\
      Options for phase modulation of silicon waveguides. a) thermal tuning with \ch{TiN} filament; b) thermal tuning with embedded photoconductive heater; c) PN/PIN junction across the waveguide for injection and/or depletion modulation; d) III-V/Si hybrid waveguide; e) metal-oxide-semiconductor (MOS), where the `metal' is actually an active semiconductor; f) lithium niobate cladding adds a strong electrooptic effect; g) 2 single-layer-graphene (SLG) `capacitor'; h) non-volatile phase change material.
      % Letters in parentheses correspond to Fig.~\ref{fig:box2:phase-modulation}.
      $^1$This bandwidth was not yet shown experimentally. A big challenge is to reduce the contact resistance with Graphene, reducing RC-loading effect.
      $^2$Not experimentally shown at high-speed.
      $^3$Demonstrated up to \SI{20}{MHz}.

      \caption{Efficiency and speed of various index modulation techniques on silicon photonics.
      }
      
      \label{tab:modulation-options}
    \end{table}

    Moving beyond pure silicon approaches, one can make hybrid waveguides consisting of a silicon core and additional materials with favorable index modulation properties that are placed close enough to the core that interact with the evanescent field. Some examples include III-V semiconductors~\cite{Chen2011}, lithium niobate~\cite{He2019}, and graphene~\cite{Sorianello2018}. These mechanisms are faster and require much less power compared to heaters, but typically provide smaller tuning range before the onset of electrical damage. Tuning methods based on chalcogenide phase change materials (PCMs) allow weights to retain their values without active power consumption after being set~\cite{Feldmann:19,Rios2015}. Examples of some non-volatile photonic weight implementations include \ch{Si3N4} integrated waveguides~\cite{Feldmann:19} and metal-sulphide fibers~\cite{Gholipour:2015}.

    \begin{figure*}[ht]
      \centering
      \includegraphics[width=\textwidth]{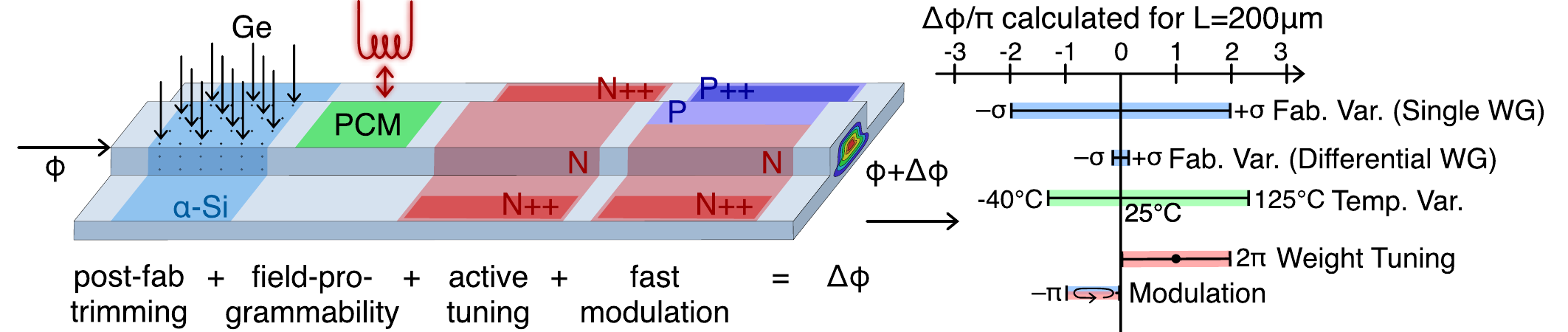}
      \caption{Active trimming and Phase modulation strategy. In this example strategy, Ge ion implantation is used as a post-fabrication trimming technique, a PCM is used as a field-programmable non-volatile memory, an N-doped heater is used for weight tuning and configuration, and a standard PN junction is used for fast modulation. The graph on the right shows the relative phase variability from the environment compared to the necessary variation required for modulation and weight configuration. These values were experimentally computed from test structures fabricated through a standard silicon photonics foundry~\cite{Chrostowski2014}.
      }
      \label{fig:box2:phase-modulation}
    \end{figure*}

    Another important variation source is ambient temperature. Automotive-class devices must have an operating temperature range of \SI{-40}{\celsius} to \SI{125}{\celsius}). Due to silicon's large thermo-optic coefficient, this temperature range results in a large resonance drift, albeit no worse than the original fabrication variation (Fig.~\ref{fig:box2:phase-modulation}). Depending on the final application of the circuit, these two sources of variation can be addressed with different index modulation options from Table~\ref{tab:modulation-options}.

    We propose two practical options for addressing variations, including the ones present in automotive temperature ranges. The first is to use post-fabrication trimming to lock resonators to their desired resonance wavelengths at a set operating temperature, which can be chosen, for example, to be above room temperature to avoid condensation. A thermoelectric controller can then be used to stabilize the chip's resting temperature and fine-tuning can be performed via either microheaters or PN junctions, depending on the required operation speed and actuation range. This greatly simplifies the photonic integrated circuit design, but the use of temperature control adds thermal engineering complexity to the packaging. Another option is to rely on active trimming, in which on-chip microheaters are used to compensate for wide operating temperature ranges. Accompanying this, fine-tuning can be performed via microheaters or PN junctions. This approach requires more control circuitry, but eases the overall burden on packaging~\cite{grillanda_non-invasive_2014,derose_silicon_2010}. Both approaches can be visualized in Fig.~\ref{fig:box2:phase-modulation}.

    \subsection{Electrical Actuation and Sensing}
    
    As reviewed above, there are multiple physical phenomena that are available for electrically adjusting the refractive index of a waveguide, and consequently the resonance wavelength of an integrated resonator. Since the weight is determined by the distance between a lightwave's wavelength and a resonator's resonance wavelength~(Table~\ref{tab:simplified_model}), we would ideally have a sensor that can directly determine resonance wavelength (\(\lambda_0\)). With such a sensor coupled with resonance actuation, we would only need to model the mapping between resonance wavelength and weight (i.e., a Lorentzian function for microring resonators) in order to precisely perform weighting. In reality, it is impractical to measure \(\lambda_0\) directly because it would require a precision spectrometer. 
    
    One practical alternative is to use a temperature sensor~\cite{Saeedi2015} surrounding the resonator. This allows absolute temperature stabilization independent from the environment's temperature. Although this provides an efficient mechanism for `fixing' a weight, it usually requires sourcing very precise currents and reading tiny voltage signals with high dynamic range. So far, it has only been successfully employed to stabilize microring resonators around maximum transmission, which is useful for optical switching but not smoothly-variable weighting.

    \begin{figure}[!ht]
    \centering
    \includegraphics[width=\linewidth]{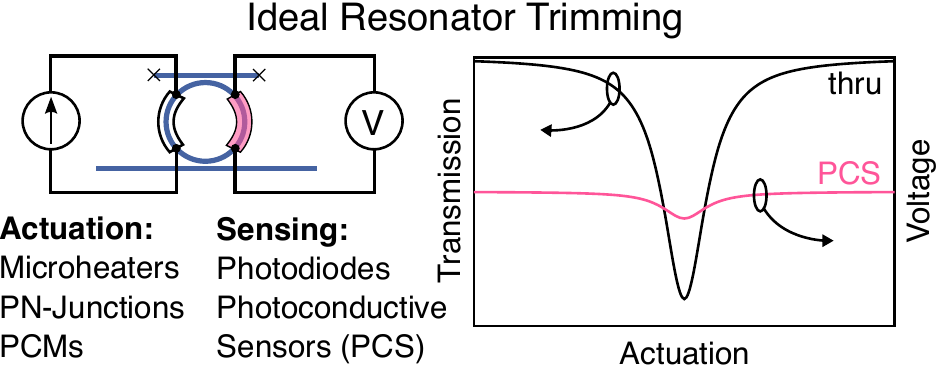}\\
    \caption{Schematic of a resonator trimming circuit. Effective trimming requires an actuation stage and a sensing stage. Dropped power reaches a maximum when the actuation signal brings the resonator's resonant wavelength to exactly the signal's wavelength.}
    \label{fig:resonator-trimming}
    \end{figure}

    Another practical alternative is to measure the optical power circulating within a resonator with a photosensitive element, shown in Fig.~\ref{fig:resonator-trimming}. This method only works if the weight value is proportional to the circulating power relative to input power, and if input power is otherwise known. A simple method for measuring circulating power is by lightly doping the resonator's waveguide, which results in a photoconductive resistor. This resistor can be used for both heating and sensing the circulating power, and was successfully employed in both resonance-locking circuits~\cite{Jayatilleka2015} and weight control~\cite{Tait:18fb}.

    This sensing modality will suffer if parasitic resistances are too large. Fortunately, when chips are packaged with low-impedance wirebonds and properly designed printed circuit boards (PCBs) the parasitic resistances become insignificant compared to the heater resistance~\cite{huang_demonstration_2020}. We have found that a resistance range that works well is \SIrange{1.5}{2.5}{\kilo Ohm}. In a typical silicon waveguide, a \textpi~phase shift is achieved with the application of \around\SI{20}{\milli\watt} of heating via the thermooptic effect. This helps limit the required voltage to under \SI{10}{\volt} and current to under \SI{4}{\mA}, within the ranges of widely available DAC and ADC circuits. Finally, we found that adding noise filtering at the PCB level and shielding cables are key to combat power supply ripple and electromagnetic interference.

    % \fxwarning{In practice, we have seen that wirebonding test chips, adding noise filtering at the PCB level, and shielding cables have all independently helped decrease noise and increase precision. In addition, any systematic error that is not corrected by calibration, in practice, should also be included in the precision calculation. However, it is often ignored when reported in papers. As a rule of thumb we find that the accuracy is usually bounded above by the precision, because a calibration algorithm is negatively impacted by noisy data, and sources of random error correlate with nonlinearities causing the systematic error. The most common source of random error are power supply ripple, electromagnetic interference on the PCB, unshielded cables, and, to a lesser extent, shot noise and thermal noise on the photodetector.}
    
    \subsection{Feedback Control}
    \label{subsec:feedback control}

    Equipped with an appropriate sensing element, it is then possible to build a closed-loop control circuit that dynamically responds to external disturbances to a resonator. Similarly, with a sensor for each resonator in a weight bank it becomes possible to correct for disturbances to each resonator individually. This modality of control based on sensed signals is called a feedback (or closed-loop) control scheme. Compared to the feedforward scheme introduced in Sec.~\ref{sec:calibration and control}, the feedback scheme offers superior accuracy and precision as well as reduced control complexity.

    \begin{figure}[!t]
      \centering
      \includegraphics[width=\linewidth]{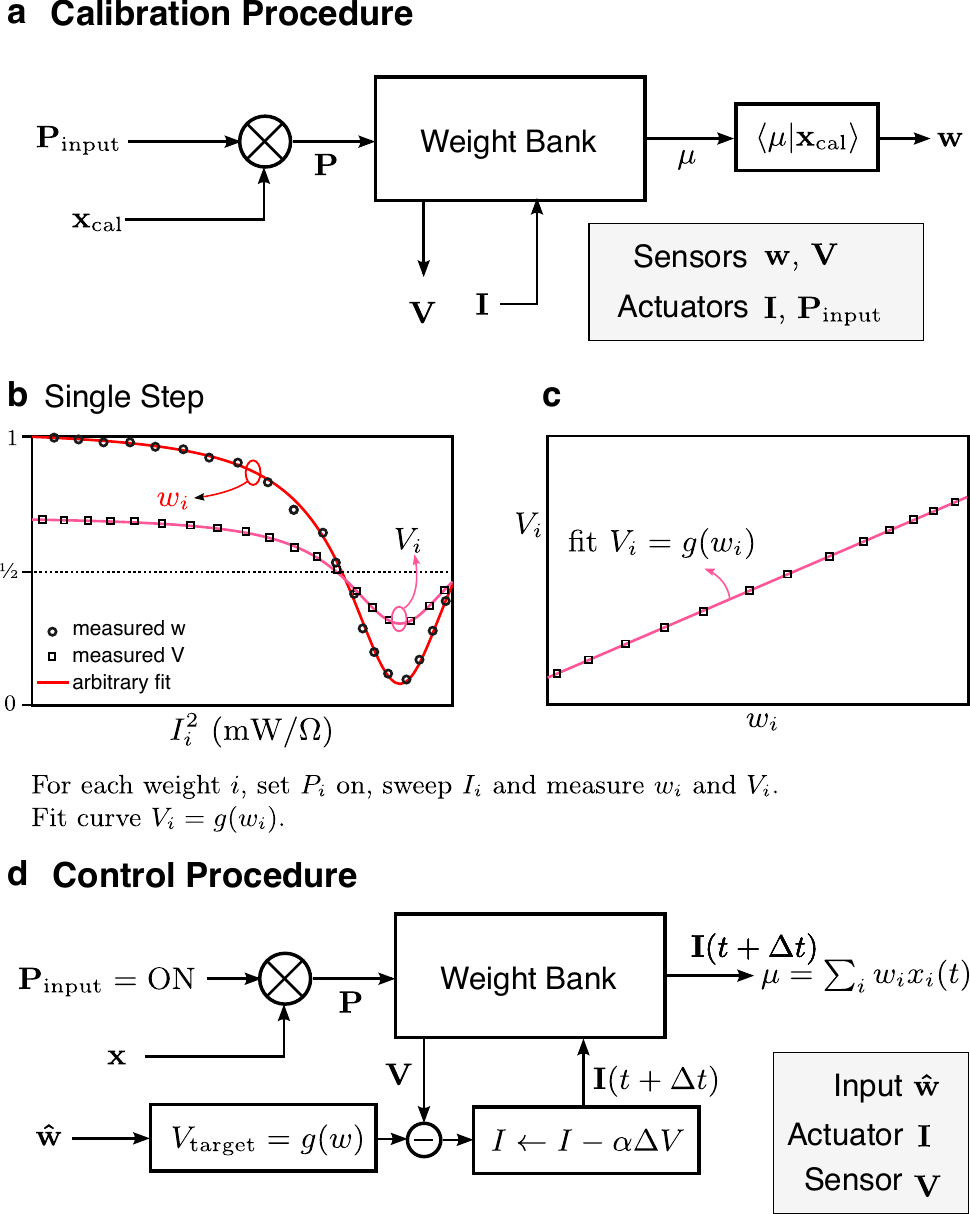}\\
      \caption{Procedure for feedback calibration and control. The calibration can be performed in a single sweep for each resonator (\(\mathcal{O}(N)\)) with a specially crafted calibration signal $x_\text{cal}$ (a). The goal is to fit the relationship between effective weight and the sensor voltage signal (b), which is ideally a linear relationship (c). Once that is established, the control procedure (d) takes a commanded weight, computes the target sensor voltage, and uses a feedback loop to update the actuation electrical current until that voltage is reached. This feedback loop should be resistant to environmental variations so long as the feedback loop can operate at a higher speed than the variations.
      % Tait's comments: On these axes, put 1) labels, 2) units, 3) bound/tick values, even if it implied that is -1 to 1, 4) some way to refer to them, such as sub-panel (a,b,c...). Consider a legend to show experimental
      }
      \label{fig:feedback calibration and control}
    \end{figure}

    A typical control circuit involves a current-based actuator, voltage-based sensors, and a model that maps the sensing signal to the desired control quantity (Fig.~\ref{fig:feedback calibration and control}). Having a sensor for every weight element simplifies the feedback control law from one large model for the weight bank to a collection of independent models for each resonator. Unlike feedforward control, it is no longer necessary to take into account the thermal crosstalk between resonators or the ambient temperature because the closed-loop circuit corrects for any deviation from the target.

    % Figure~\ref{fig:resonator-trimming} illustrates an example of a microring resonator weight designed for feedback control. 
    A photoconductive sensor (PCS) provides a voltage signal that is linearly dependent to the transmission function. The transmission function can be translated in wavelength space with an actuator, leading to smoothly-controllable weight values. Rather than modeling how transmission is affected by actuation and other environmental variables, the controller can sense the voltage directly and simply adjust the actuation accordingly.

    Despite using a closed-loop/feedback control circuit, each resonator may vary slightly due to fabrication imperfections. In order to compensate for this, all imperfections must be parameterized and calibrated out on a chip-by-chip basis.
    Fortunately, this model only has to be built once per device. Therefore, it can be done at the manufacturing facility and stored in a read-only memory co-packaged with the chip.
    % See  Sec.~\ref{sec:calibration and control} for a discussion of the complexity of calibration procedures and the accuracy of control circuits.
    
    \section{Conclusion}
    
    Practical neuromorphic photonic processors employing resonator-based weight banks must be able to achieve high precision and accuracy while being readily manufacturable and deployable in variable/dynamic environments. This is made particularly challenging by the fact that resonators are highly sensitive to both manufacturing variations and environmental disturbances. In order to compensate for this, individual resonator weights must be specifically designed with calibration and control in mind. 
    
    In this paper we have outlined index tuning and sensing mechanisms that can be used for actuation and sensing of weights, as well as feedforward and feedback calibration/control methodologies for use with weight banks. With multiple index tuning mechanisms with varying speed, efficiency, and dynamic range available, system designers can choose which mechanism or combination of mechanisms are suitable based on application needs.
    
    As an example, we outline a fabrication design strategy that can compensate for worst-case fabrication variation ($\Delta\phi=\pm2\pi$) and employment in automotive-class operating temperature ranges (\SI{-40}{\celsius} to \SI{125}{\celsius}). Germanium ion implantation is used for post-fabrication trimming, a phase-change material (PCM) is used to compensate for large temperature variations, an N-doped heater is used for lower-bandwidth weight tuning, and a PN junction is used for fast modulation. Complementing this, photoconductive sensors are used on each resonator in order to enable feedback calibration/control procedures.
    
    % Key takeaways:
    % \begin{enumerate}
    %     \item Individual resonator weights must be designed for calibration and control on a weight bank setting.
    %     \item Resonator weights are particularly susceptible to environmental disturbances which should be controlled in real time. (feedforward and feedback)
    %     \item This control procedure must involve several index tuning mechanisms to achieve both accuracy and dynamic range targets.
    % \end{enumerate}

% Can use something like this to put references on a page
% by themselves when using endfloat and the captionsoff option.
% \newpage

% \fxwarning{What is this doing here? \cite{bandyopadhyay_hardware_2021} --tlima}

% trigger a \newpage just before the given reference
% number - used to balance the columns on the last page
% adjust value as needed - may need to be readjusted if
% the document is modified later
%\IEEEtriggeratref{8}
% The "triggered" command can be changed if desired:
%\IEEEtriggercmd{\enlargethispage{-5in}}

% \appendices

\section{Appendix: Microring Resonator Weight Design Methodology}
\label{sec:designTool}

In this section, we outline our design methodology for symmetric add-drop microring resonators. We follow the nomenclature from Ref.~\cite{bogaerts_silicon_2012}. The ring's \emph{add} and \emph{drop} transmission coefficients can be computed analytically using transfer matrices with only four parameters: self-coupling coefficient $r$ (thru port amplitude),  round-trip loss coefficient $a \rightarrow ra$ (accounting for the second coupler's coupling loss), cavity length $L$, and effective index of refraction $n_{\text{eff}}$ as a function of wavelength. Each parameter depends on the geometry of the resonator such as waveguide width, the gap between the bus waveguide and the ring, the ring's radius $R$, and the doping level of the waveguide. $r$ can be computed through finite-difference time domain (FDTD) calculations of the coupler section. $a$ is related to power attenuation coefficient $\alpha$ via
\(
a^2 = \exp{\left( -\alpha L \right)}
\)
, and $\alpha$ is determined by the combination of the loss mechanisms in the ring. This includes the propagation loss from absorption and waveguide imperfections (process-dependent) and bending loss. $L = 2\pi R$ for ring resonators, and the waveguide effective index depends on waveguide material and geometry and is calculated from eigenmode simulations. From the resulting transmission coefficients, the extinction ratios -- corresponding to the ratio between maximum and minimum weights -- can be computed by evaluating the ratio between the maximum and minimum transmission values on either the \emph{thru} and \emph{drop} ports.

The quantities of interest for weighting can be extracted from the parameters above. The free spectral range is calculated as
$$ \text{FSR}(\lambda) = \frac{\lambda^2}{2\pi R}\left( n_{\text{eff}}(\lambda) - \lambda\frac{\partial n_{\text{eff}}}{\partial \lambda} \right)^{-1}. $$
The finesse determines roughly how many WDM channels (or resonators) can share the same bus waveguide $$\mathcal{F} = \pi\arccos\left(\frac{2ar^2}{1+a^2r^4}\right).$$
The Q-factor approximates the optical rise and fall times of the resonator:
$$ Q = \frac{\lambda\mathcal{F}}{\text{FSR}}. $$ 
Finally, the \emph{thru} and \emph{drop} extinction ratios allow calculation of the weighting dynamic range (assuming the index tuning mechanism can be used to tune the resonator on and off-resonance):
$$ \text{ER}_{\text{thru}} = \frac{(a+1)^2}{(a-1)^2}\frac{(1-r^2a)^2}{(1+r^2a)^2},
\text{ER}_{\text{drop}} = \frac{(1 + r^2a)^2}{(1 - r^2a)^2}. $$
% and insertion loss 
% $$ IL = \frac{r^2(1+a)^2}{(1+r^2a)^2}$$ 

We used Lumerical FDTD to perform the finite-difference time domain coupler simulations to extract $r(\lambda)$ as a function of geometry, and Lumerical MODE for the curved waveguide eigensolver simulations to obtain $n_{\text{eff}}(\lambda)$. A Python script implements the above analysis pipeline, and the results can be visualized in a Jupyter Notebook. The Python script and the results of the Lumerical simulations are made accessible online\footnote{\url{https://github.com/lightwave-lab/microring-resonator-weight-design}}.

\begin{thebibliography}{10}
\providecommand{\url}[1]{#1}
\csname url@samestyle\endcsname
\providecommand{\newblock}{\relax}
\providecommand{\bibinfo}[2]{#2}
\providecommand{\BIBentrySTDinterwordspacing}{\spaceskip=0pt\relax}
\providecommand{\BIBentryALTinterwordstretchfactor}{4}
\providecommand{\BIBentryALTinterwordspacing}{\spaceskip=\fontdimen2\font plus
\BIBentryALTinterwordstretchfactor\fontdimen3\font minus
  \fontdimen4\font\relax}
\providecommand{\BIBforeignlanguage}[2]{{%
\expandafter\ifx\csname l@#1\endcsname\relax
\typeout{** WARNING: IEEEtran.bst: No hyphenation pattern has been}%
\typeout{** loaded for the language `#1'. Using the pattern for}%
\typeout{** the default language instead.}%
\else
\language=\csname l@#1\endcsname
\fi
#2}}
\providecommand{\BIBdecl}{\relax}
\BIBdecl

\bibitem{horowitz_computings_2014}
\BIBentryALTinterwordspacing
M.~Horowitz, ``\BIBforeignlanguage{en}{Computing's energy problem (and what we
  can do about it)},'' in \emph{\BIBforeignlanguage{en}{2014 {IEEE}
  {International} {Solid}-{State} {Circuits} {Conference} {Digest} of
  {Technical} {Papers} ({ISSCC})}}.\hskip 1em plus 0.5em minus 0.4em\relax San
  Francisco, CA, USA: IEEE, Feb. 2014, pp. 10--14, 310 citations (Crossref)
  [2021-12-04]. [Online]. Available:
  \url{http://ieeexplore.ieee.org/document/6757323/}
\BIBentrySTDinterwordspacing

\bibitem{raina_large-scale_2009}
\BIBentryALTinterwordspacing
R.~Raina, A.~Madhavan, and A.~Y. Ng, ``\BIBforeignlanguage{en}{Large-scale deep
  unsupervised learning using graphics processors},'' in
  \emph{\BIBforeignlanguage{en}{Proceedings of the 26th {Annual}
  {International} {Conference} on {Machine} {Learning} - {ICML} '09}}.\hskip
  1em plus 0.5em minus 0.4em\relax Montreal, Quebec, Canada: ACM Press, 2009,
  pp. 1--8, 252 citations (Crossref) [2021-12-04]. [Online]. Available:
  \url{http://portal.acm.org/citation.cfm?doid=1553374.1553486}
\BIBentrySTDinterwordspacing

\bibitem{jouppi_-datacenter_2017}
\BIBentryALTinterwordspacing
N.~P. Jouppi, C.~Young, N.~Patil, D.~Patterson, G.~Agrawal, R.~Bajwa, S.~Bates,
  S.~Bhatia, N.~Boden, A.~Borchers, R.~Boyle, P.-l. Cantin, C.~Chao, C.~Clark,
  J.~Coriell, M.~Daley, M.~Dau, J.~Dean, B.~Gelb, T.~V. Ghaemmaghami,
  R.~Gottipati, W.~Gulland, R.~Hagmann, C.~R. Ho, D.~Hogberg, J.~Hu, R.~Hundt,
  D.~Hurt, J.~Ibarz, A.~Jaffey, A.~Jaworski, A.~Kaplan, H.~Khaitan,
  D.~Killebrew, A.~Koch, N.~Kumar, S.~Lacy, J.~Laudon, J.~Law, D.~Le, C.~Leary,
  Z.~Liu, K.~Lucke, A.~Lundin, G.~MacKean, A.~Maggiore, M.~Mahony, K.~Miller,
  R.~Nagarajan, R.~Narayanaswami, R.~Ni, K.~Nix, T.~Norrie, M.~Omernick,
  N.~Penukonda, A.~Phelps, J.~Ross, M.~Ross, A.~Salek, E.~Samadiani, C.~Severn,
  G.~Sizikov, M.~Snelham, J.~Souter, D.~Steinberg, A.~Swing, M.~Tan,
  G.~Thorson, B.~Tian, H.~Toma, E.~Tuttle, V.~Vasudevan, R.~Walter, W.~Wang,
  E.~Wilcox, and D.~H. Yoon, ``\BIBforeignlanguage{en}{In-{Datacenter}
  {Performance} {Analysis} of a {Tensor} {Processing} {Unit}},'' in
  \emph{\BIBforeignlanguage{en}{Proceedings of the 44th {Annual}
  {International} {Symposium} on {Computer} {Architecture}}}.\hskip 1em plus
  0.5em minus 0.4em\relax Toronto ON Canada: ACM, Jun. 2017, pp. 1--12.
  [Online]. Available: \url{https://dl.acm.org/doi/10.1145/3079856.3080246}
\BIBentrySTDinterwordspacing

\bibitem{davies_loihi_2018}
\BIBentryALTinterwordspacing
M.~Davies, N.~Srinivasa, T.-H. Lin, G.~Chinya, Y.~Cao, S.~H. Choday, G.~Dimou,
  P.~Joshi, N.~Imam, S.~Jain, Y.~Liao, C.-K. Lin, A.~Lines, R.~Liu,
  D.~Mathaikutty, S.~McCoy, A.~Paul, J.~Tse, G.~Venkataramanan, Y.-H. Weng,
  A.~Wild, Y.~Yang, and H.~Wang, ``Loihi: {A} {Neuromorphic} {Manycore}
  {Processor} with {On}-{Chip} {Learning},'' \emph{IEEE Micro}, vol.~38, no.~1,
  pp. 82--99, Jan. 2018, 839 citations (Crossref) [2021-12-04]. [Online].
  Available: \url{http://ieeexplore.ieee.org/document/8259423/}
\BIBentrySTDinterwordspacing

\bibitem{akopyan_truenorth_2015}
\BIBentryALTinterwordspacing
F.~Akopyan, J.~Sawada, A.~Cassidy, R.~Alvarez-Icaza, J.~Arthur, P.~Merolla,
  N.~Imam, Y.~Nakamura, P.~Datta, G.-J. Nam, B.~Taba, M.~Beakes, B.~Brezzo,
  J.~B. Kuang, R.~Manohar, W.~P. Risk, B.~Jackson, and D.~S. Modha,
  ``\BIBforeignlanguage{en}{{TrueNorth}: {Design} and {Tool} {Flow} of a 65
  {mW} 1 {Million} {Neuron} {Programmable} {Neurosynaptic} {Chip}},''
  \emph{\BIBforeignlanguage{en}{IEEE Transactions on Computer-Aided Design of
  Integrated Circuits and Systems}}, vol.~34, no.~10, pp. 1537--1557, Oct.
  2015, 470 citations (Crossref) [2021-12-04]. [Online]. Available:
  \url{http://ieeexplore.ieee.org/document/7229264/}
\BIBentrySTDinterwordspacing

\bibitem{Merolla2014}
\BIBentryALTinterwordspacing
P.~A. Merolla, J.~V. Arthur, R.~Alvarez-Icaza, A.~S. Cassidy, J.~Sawada,
  F.~Akopyan, B.~L. Jackson, N.~Imam, C.~Guo, Y.~Nakamura, B.~Brezzo, I.~Vo,
  S.~K. Esser, R.~Appuswamy, B.~Taba, A.~Amir, M.~D. Flickner, W.~P. Risk,
  R.~Manohar, and D.~S. Modha, ``A million spiking-neuron integrated circuit
  with a scalable communication network and interface,'' \emph{Science}, vol.
  345, no. 6197, pp. 668--673, 2014, 1771 citations (Crossref) [2021-12-04]
  ISBN: 1853467960. [Online]. Available:
  \url{http://www.sciencemag.org/cgi/doi/10.1126/science.1254642}
\BIBentrySTDinterwordspacing

\bibitem{mead_neuromorphic_1990}
\BIBentryALTinterwordspacing
C.~Mead, ``\BIBforeignlanguage{en}{Neuromorphic electronic systems},''
  \emph{\BIBforeignlanguage{en}{Proceedings of the IEEE}}, vol.~78, no.~10, pp.
  1629--1636, Oct. 1990, 1012 citations (Crossref) [2021-12-04]. [Online].
  Available: \url{http://ieeexplore.ieee.org/document/58356/}
\BIBentrySTDinterwordspacing

\bibitem{prucnal_neuromorphic_2017}
\BIBentryALTinterwordspacing
P.~R. Prucnal and B.~J. Shastri, ``\BIBforeignlanguage{English}{Neuromorphic
  {Photonics}.}'' 2017, archive: /z-wcorg/ ISBN: 9781498725248 1498725244
  9781351987615 1351987615 9781351987622 1351987623 9781351987608 1351987607
  9781315370590 131537059X Library Catalog: http://worldcat.org Place: Boca
  Raton Publisher: CRC Press. [Online]. Available:
  \url{http://www.crcnetbase.com/isbn/}
\BIBentrySTDinterwordspacing

\bibitem{Tait2016}
\BIBentryALTinterwordspacing
A.~N. Tait, A.~X. Wu, T.~F. de~Lima, E.~Zhou, B.~J. Shastri, M.~A. Nahmias, and
  P.~R. Prucnal, ``Microring {Weight} {Banks},'' \emph{IEEE Journal of Selected
  Topics in Quantum Electronics}, vol.~22, no.~6, pp. 312--325, Nov. 2016, 50
  citations (Crossref) [2021-10-26] 55 citations (Semantic Scholar/DOI)
  [2021-09-15]. [Online]. Available:
  \url{http://ieeexplore.ieee.org/lpdocs/epic03/wrapper.htm?arnumber=7479545}
\BIBentrySTDinterwordspacing

\bibitem{ferreira_de_lima_machine_2019}
\BIBentryALTinterwordspacing
T.~Ferreira~de Lima, H.-T.~T. Peng, A.~N. Tait, M.~A. Nahmias, H.~B. Miller,
  B.~J. Shastri, P.~R. Prucnal, T.~F. De~Lima, H.-T.~T. Peng, A.~N. Tait, M.~A.
  Nahmias, H.~B. Miller, B.~J. Shastri, and P.~R. Prucnal, ``Machine {Learning}
  with {Neuromorphic} {Photonics},'' \emph{Journal of Lightwave Technology},
  vol.~37, no.~5, pp. 1515--1534, Mar. 2019. [Online]. Available:
  \url{https://ieeexplore.ieee.org/document/8662590/}
\BIBentrySTDinterwordspacing

\bibitem{shastri_photonics_2021}
\BIBentryALTinterwordspacing
B.~J. Shastri, A.~N. Tait, T.~Ferreira~de Lima, W.~H.~P. Pernice, H.~Bhaskaran,
  C.~D. Wright, and P.~R. Prucnal, ``\BIBforeignlanguage{en}{Photonics for
  artificial intelligence and neuromorphic computing},''
  \emph{\BIBforeignlanguage{en}{Nature Photonics}}, vol.~15, no.~2, pp.
  102--114, Feb. 2021, 94 citations (Crossref) [2022-01-27] Number: 2
  Publisher: Nature Publishing Group. [Online]. Available:
  \url{https://www.nature.com/articles/s41566-020-00754-y].}
\BIBentrySTDinterwordspacing

\bibitem{huang_silicon_2021}
\BIBentryALTinterwordspacing
C.~Huang, S.~Fujisawa, T.~F. de~Lima, A.~N. Tait, E.~C. Blow, Y.~Tian,
  S.~Bilodeau, A.~Jha, F.~Yaman, H.-T. Peng, H.~G. Batshon, B.~J. Shastri,
  Y.~Inada, T.~Wang, and P.~R. Prucnal, ``\BIBforeignlanguage{en}{A silicon
  photonic–electronic neural network for fibre nonlinearity compensation},''
  \emph{\BIBforeignlanguage{en}{Nature Electronics}}, vol.~4, no.~11, pp.
  837--844, Nov. 2021, 0 citations (Crossref) [2022-01-27] Bandiera\_abtest: a
  Cg\_type: Nature Research Journals Number: 11 Primary\_atype: Research
  Publisher: Nature Publishing Group Subject\_term: Fibre optics and optical
  communications;Integrated optics;Optoelectronic devices and components
  Subject\_term\_id:
  fibre-optics-and-optical-communications;integrated-optics;optoelectronic-devices-and-components.
  [Online]. Available: \url{https://www.nature.com/articles/s41928-021-00661-2}
\BIBentrySTDinterwordspacing

\bibitem{frantz_digital_2000}
\BIBentryALTinterwordspacing
G.~Frantz, ``\BIBforeignlanguage{en}{Digital signal processor trends},''
  \emph{\BIBforeignlanguage{en}{IEEE Micro}}, vol.~20, no.~6, pp. 52--59, Dec.
  2000, 65 citations (Crossref) [2021-12-04]. [Online]. Available:
  \url{http://ieeexplore.ieee.org/document/888703/}
\BIBentrySTDinterwordspacing

\bibitem{ramacher_design_1991}
\BIBentryALTinterwordspacing
U.~Ramacher, J.~Beichter, W.~Raab, J.~Anlauf, N.~Brüls, U.~Hachmann, and
  M.~Wesseling, ``Design of a 1st {Generation} {Neurocomputer},'' in
  \emph{{VLSI} {Design} of {Neural} {Networks}}, U.~Ramacher and U.~Rückert,
  Eds.\hskip 1em plus 0.5em minus 0.4em\relax Boston, MA: Springer US, 1991,
  pp. 271--310. [Online]. Available:
  \url{http://link.springer.com/10.1007/978-1-4615-3994-0_14}
\BIBentrySTDinterwordspacing

\bibitem{xu_cascaded_2006}
\BIBentryALTinterwordspacing
Q.~Xu, B.~Schmidt, J.~Shakya, and M.~Lipson, ``\BIBforeignlanguage{en}{Cascaded
  silicon micro-ring modulators for {WDM} optical interconnection},''
  \emph{\BIBforeignlanguage{en}{Optics Express}}, vol.~14, no.~20, p. 9431,
  2006, 190 citations (Crossref) [2021-12-02]. [Online]. Available:
  \url{https://www.osapublishing.org/oe/abstract.cfm?uri=oe-14-20-9431}
\BIBentrySTDinterwordspacing

\bibitem{rahim_open-access_2018}
\BIBentryALTinterwordspacing
A.~Rahim, T.~Spuesens, R.~Baets, and W.~Bogaerts,
  ``\BIBforeignlanguage{en}{Open-{Access} {Silicon} {Photonics}: {Current}
  {Status} and {Emerging} {Initiatives}},''
  \emph{\BIBforeignlanguage{en}{Proceedings of the IEEE}}, vol. 106, no.~12,
  pp. 2313--2330, Dec. 2018. [Online]. Available:
  \url{https://ieeexplore.ieee.org/document/8540508/}
\BIBentrySTDinterwordspacing

\bibitem{Giewont2019}
\BIBentryALTinterwordspacing
K.~Giewont, S.~Hu, B.~Peng, M.~Rakowski, S.~Rauch, J.~C. Rosenberg, A.~Sahin,
  I.~Stobert, A.~Stricker, K.~Nummy, F.~A. Anderson, J.~Ayala, T.~Barwicz,
  Y.~Bian, K.~K. Dezfulian, D.~M. Gill, and T.~Houghton, ``300-mm {Monolithic}
  {Silicon} {Photonics} {Foundry} {Technology},'' \emph{IEEE Journal of
  Selected Topics in Quantum Electronics}, vol.~25, no.~5, pp. 1--11, Sep.
  2019, 70 citations (Crossref) [2021-12-02] Publisher: IEEE. [Online].
  Available: \url{https://ieeexplore.ieee.org/document/8678809/}
\BIBentrySTDinterwordspacing

\bibitem{chrostowski_silicon_2019}
\BIBentryALTinterwordspacing
L.~Chrostowski, H.~Shoman, M.~Hammood, H.~Yun, J.~Jhoja, E.~Luan, S.~Lin,
  A.~Mistry, D.~Witt, N.~A.~F. Jaeger, S.~Shekhar, H.~Jayatilleka, P.~Jean,
  S.~B.-d. Villers, J.~Cauchon, W.~Shi, C.~Horvath, J.~N. Westwood-Bachman,
  K.~Setzer, M.~Aktary, N.~S. Patrick, R.~J. Bojko, A.~Khavasi, X.~Wang,
  T.~Ferreira~de Lima, A.~N. Tait, P.~R. Prucnal, D.~E. Hagan, D.~Stevanovic,
  and A.~P. Knights, ``\BIBforeignlanguage{en}{Silicon {Photonic} {Circuit}
  {Design} {Using} {Rapid} {Prototyping} {Foundry} {Process} {Design}
  {Kits}},'' \emph{\BIBforeignlanguage{en}{IEEE Journal of Selected Topics in
  Quantum Electronics}}, vol.~25, no.~5, pp. 1--26, Sep. 2019, 34 citations
  (Crossref) [2021-12-02] 35 citations (Semantic Scholar/DOI) [2021-09-15].
  [Online]. Available: \url{https://ieeexplore.ieee.org/document/8718393/}
\BIBentrySTDinterwordspacing

\bibitem{padmaraju_resolving_2014}
\BIBentryALTinterwordspacing
K.~Padmaraju and K.~Bergman, ``\BIBforeignlanguage{en}{Resolving the thermal
  challenges for silicon microring resonator devices},''
  \emph{\BIBforeignlanguage{en}{Nanophotonics}}, vol.~3, no. 4-5, pp. 269--281,
  Aug. 2014. [Online]. Available:
  \url{https://www.degruyter.com/document/doi/10.1515/nanoph-2013-0013/html}
\BIBentrySTDinterwordspacing

\bibitem{derose_silicon_2010}
\BIBentryALTinterwordspacing
C.~T. DeRose, M.~R. Watts, D.~C. Trotter, D.~L. Luck, G.~N. Nielson, and R.~W.
  Young, ``\BIBforeignlanguage{en}{Silicon {Microring} {Modulator} with
  {Integrated} {Heater} and {Temperature} {Sensor} for {Thermal} {Control}},''
  in \emph{\BIBforeignlanguage{en}{Conference on {Lasers} and
  {Electro}-{Optics} 2010}}.\hskip 1em plus 0.5em minus 0.4em\relax San Jose,
  California: OSA, 2010, p. CThJ3, 21 citations (Crossref) [2021-12-02].
  [Online]. Available:
  \url{https://www.osapublishing.org/abstract.cfm?URI=CLEO-2010-CThJ3}
\BIBentrySTDinterwordspacing

\bibitem{grillanda_non-invasive_2014}
\BIBentryALTinterwordspacing
S.~Grillanda, M.~Carminati, F.~Morichetti, P.~Ciccarella, A.~Annoni,
  G.~Ferrari, M.~Strain, M.~Sorel, M.~Sampietro, and A.~Melloni,
  ``\BIBforeignlanguage{en}{Non-invasive monitoring and control in silicon
  photonics using {CMOS} integrated electronics},''
  \emph{\BIBforeignlanguage{en}{Optica}}, vol.~1, no.~3, p. 129, Sep. 2014.
  [Online]. Available:
  \url{https://www.osapublishing.org/abstract.cfm?URI=optica-1-3-129}
\BIBentrySTDinterwordspacing

\bibitem{bogaerts_silicon_2012}
\BIBentryALTinterwordspacing
W.~Bogaerts, P.~De~Heyn, T.~Van~Vaerenbergh, K.~De~Vos, S.~Kumar~Selvaraja,
  T.~Claes, P.~Dumon, P.~Bienstman, D.~Van~Thourhout, and R.~Baets,
  ``\BIBforeignlanguage{en}{Silicon microring resonators},''
  \emph{\BIBforeignlanguage{en}{Laser \& Photonics Reviews}}, vol.~6, no.~1,
  pp. 47--73, 2012, 1261 citations (Crossref) [2021-12-04] \_eprint:
  https://onlinelibrary.wiley.com/doi/pdf/10.1002/lpor.201100017. [Online].
  Available:
  \url{https://onlinelibrary.wiley.com/doi/abs/10.1002/lpor.201100017}
\BIBentrySTDinterwordspacing

\bibitem{eid2016fsr}
N.~Eid, R.~Boeck, H.~Jayatilleka, L.~Chrostowski, W.~Shi, and N.~A. Jaeger,
  ``Fsr-free silicon-on-insulator microring resonator based filter with bent
  contra-directional couplers,'' \emph{Optics express}, vol.~24, no.~25, pp.
  29\,009--29\,021, 2016.

\bibitem{zhou_compact_2017}
\BIBentryALTinterwordspacing
H.~Zhou, C.~Qiu, X.~Jiang, Q.~Zhu, Y.~He, Y.~Zhang, Y.~Su, and R.~Soref,
  ``\BIBforeignlanguage{en}{Compact, submilliwatt, 2 × 2 silicon thermo-optic
  switch based on photonic crystal nanobeam cavities},''
  \emph{\BIBforeignlanguage{en}{Photonics Research}}, vol.~5, no.~2, p. 108,
  Apr. 2017, 33 citations (Crossref) [2022-01-12]. [Online]. Available:
  \url{https://www.osapublishing.org/abstract.cfm?URI=prj-5-2-108}
\BIBentrySTDinterwordspacing

\bibitem{nedeljkovic_free-carrier_2011}
M.~Nedeljkovic, R.~Soref, and G.~Z. Mashanovich, ``Free-{Carrier}
  {Electrorefraction} and {Electroabsorption} {Modulation} {Predictions} for
  {Silicon} {Over} the 1–14- \${\textbackslash}mu{\textbackslash}hboxm\$
  {Infrared} {Wavelength} {Range},'' \emph{IEEE Photonics Journal}, vol.~3,
  no.~6, pp. 1171--1180, Dec. 2011, 138 citations (Crossref) [2021-12-13].

\bibitem{tait_application_2017}
A.~N. Tait, T.~Ferreira~de Lima, M.~P. Chang, M.~A. Nahmias, B.~J. Shastri, and
  P.~R. Prucnal, ``Application regime and distortion metric for multivariate
  {RF} photonics,'' in \emph{2017 {IEEE} {Optical} {Interconnects} {Conference}
  ({OI})}, Jun. 2017, pp. 25--26, 2 citations (Crossref) [2021-12-04].

\bibitem{demirkiran_electro-photonic_2021}
\BIBentryALTinterwordspacing
C.~Demirkiran, F.~Eris, G.~Wang, J.~Elmhurst, N.~Moore, N.~C. Harris,
  A.~Basumallik, V.~J. Reddi, A.~Joshi, and D.~Bunandar,
  ``\BIBforeignlanguage{en}{An {Electro}-{Photonic} {System} for {Accelerating}
  {Deep} {Neural} {Networks}},'' \emph{\BIBforeignlanguage{en}{arXiv:2109.01126
  [cs]}}, Sep. 2021. [Online]. Available: \url{http://arxiv.org/abs/2109.01126}
\BIBentrySTDinterwordspacing

\bibitem{harris_linear_2018}
\BIBentryALTinterwordspacing
N.~C. Harris, J.~Carolan, D.~Bunandar, M.~Prabhu, M.~Hochberg, T.~Baehr-Jones,
  M.~L. Fanto, A.~M. Smith, C.~C. Tison, P.~M. Alsing, and D.~Englund,
  ``\BIBforeignlanguage{EN}{Linear programmable nanophotonic processors},''
  \emph{\BIBforeignlanguage{EN}{Optica}}, vol.~5, no.~12, pp. 1623--1631, Dec.
  2018, 115 citations (Crossref) [2022-01-27] Publisher: Optical Society of
  America. [Online]. Available:
  \url{https://www.osapublishing.org/optica/abstract.cfm?uri=optica-5-12-1623}
\BIBentrySTDinterwordspacing

\bibitem{williams_effects_1994}
\BIBentryALTinterwordspacing
K.~Williams, R.~Esman, and M.~Dagenais, ``\BIBforeignlanguage{en}{Effects of
  high space-charge fields on the response of microwave photodetectors},''
  \emph{\BIBforeignlanguage{en}{IEEE Photonics Technology Letters}}, vol.~6,
  no.~5, pp. 639--641, May 1994, 98 citations (Crossref) [2021-12-02].
  [Online]. Available: \url{http://ieeexplore.ieee.org/document/285565/}
\BIBentrySTDinterwordspacing

\bibitem{tait_multi-channel_2016}
\BIBentryALTinterwordspacing
A.~N. Tait, T.~F. de~Lima, M.~A. Nahmias, B.~J. Shastri, and P.~R. Prucnal,
  ``Multi-channel control for microring weight banks,'' \emph{Optics Express},
  vol.~24, no.~8, p. 8895, Apr. 2016. [Online]. Available:
  \url{https://www.osapublishing.org/abstract.cfm?URI=oe-24-8-8895}
\BIBentrySTDinterwordspacing

\bibitem{selvaraja_subnanometer_2010}
\BIBentryALTinterwordspacing
S.~K. Selvaraja, W.~Bogaerts, P.~Dumon, D.~Van~Thourhout, and R.~Baets,
  ``\BIBforeignlanguage{en}{Subnanometer {Linewidth} {Uniformity} in {Silicon}
  {Nanophotonic} {Waveguide} {Devices} {Using} {CMOS} {Fabrication}
  {Technology}},'' \emph{\BIBforeignlanguage{en}{IEEE Journal of Selected
  Topics in Quantum Electronics}}, vol.~16, no.~1, pp. 316--324, Jan. 2010.
  [Online]. Available: \url{https://ieeexplore.ieee.org/document/5325789/}
\BIBentrySTDinterwordspacing

\bibitem{chrostowski_impact_2014}
\BIBentryALTinterwordspacing
L.~Chrostowski, X.~Wang, J.~Flueckiger, Y.~Wu, Y.~Wang, and S.~T. Fard,
  ``\BIBforeignlanguage{en}{Impact of {Fabrication} {Non}-{Uniformity} on
  {Chip}-{Scale} {Silicon} {Photonic} {Integrated} {Circuits}},'' in
  \emph{\BIBforeignlanguage{en}{Optical {Fiber} {Communication}
  {Conference}}}.\hskip 1em plus 0.5em minus 0.4em\relax San Francisco,
  California: OSA, 2014, p. Th2A.37. [Online]. Available:
  \url{https://www.osapublishing.org/abstract.cfm?URI=OFC-2014-Th2A.37}
\BIBentrySTDinterwordspacing

\bibitem{lu_performance_2017}
\BIBentryALTinterwordspacing
Z.~Lu, J.~Jhoja, J.~Klein, X.~Wang, A.~Liu, J.~Flueckiger, J.~Pond, and
  L.~Chrostowski, ``\BIBforeignlanguage{EN}{Performance prediction for silicon
  photonics integrated circuits with layout-dependent correlated manufacturing
  variability},'' \emph{\BIBforeignlanguage{EN}{Optics Express}}, vol.~25,
  no.~9, pp. 9712--9733, May 2017. [Online]. Available:
  \url{https://www.osapublishing.org/oe/abstract.cfm?uri=oe-25-9-9712}
\BIBentrySTDinterwordspacing

\bibitem{Jacques2019}
M.~Jacques, A.~Samani, E.~El-Fiky, D.~Patel, Z.~Xing, and D.~V. Plant,
  ``{Optimization of thermo-optic phase-shifter design and mitigation of
  thermal crosstalk on the SOI platform},'' \emph{Optics Express}, vol.~27,
  no.~8, p. 10456, 2019.

\bibitem{Jayatilleka2015}
\BIBentryALTinterwordspacing
H.~Jayatilleka, K.~Murray, M.~{\'{A}}. Guill{\'{e}}n-Torres, M.~Caverley,
  R.~Hu, N.~A.~F. Jaeger, L.~Chrostowski, and S.~Shekhar, ``{Wavelength tuning
  and stabilization of microring-based filters using silicon in-resonator
  photoconductive heaters},'' \emph{Optics Express}, vol.~23, no.~19, p. 25084,
  sep 2015. [Online]. Available:
  \url{https://www.osapublishing.org/abstract.cfm?URI=oe-23-19-25084}
\BIBentrySTDinterwordspacing

\bibitem{Tait:18fb}
\BIBentryALTinterwordspacing
A.~N. Tait, H.~Jayatilleka, T.~{Ferreira De Lima}, P.~Y. Ma, M.~A. Nahmias,
  B.~J. Shastri, S.~Shekhar, L.~Chrostowski, and P.~R. Prucnal, ``Feedback
  control for microring weight banks,'' \emph{Opt. Express}, vol.~26, no.~20,
  pp. 26\,422--26\,443, Oct 2018. [Online]. Available:
  \url{http://www.opticsexpress.org/abstract.cfm?URI=oe-26-20- 26422}
\BIBentrySTDinterwordspacing

\bibitem{Patel2015}
D.~Patel, S.~Ghosh, M.~Chagnon, A.~Samani, V.~Veerasubramanian, M.~Osman, and
  D.~V. Plant, ``{Design, analysis, and transmission system performance of a 41
  GHz silicon photonic modulator},'' \emph{Optics Express}, vol.~23, no.~11, p.
  14263, 2015.

\bibitem{Sorianello2018}
\BIBentryALTinterwordspacing
V.~Sorianello, M.~Midrio, G.~Contestabile, I.~Asselberghs, J.~{Van Campenhout},
  C.~Huyghebaert, I.~Goykhman, A.~K. Ott, A.~C. Ferrari, and M.~Romagnoli,
  ``{Graphene-silicon phase modulators with gigahertz bandwidth},''
  \emph{Nature Photonics}, vol.~12, no.~1, pp. 40--44, 2018. [Online].
  Available: \url{http://dx.doi.org/10.1038/s41566-017-0071-6}
\BIBentrySTDinterwordspacing

\bibitem{He2019}
\BIBentryALTinterwordspacing
M.~He, M.~Xu, Y.~Ren, J.~Jian, Z.~Ruan, Y.~Xu, S.~Gao, S.~Sun, X.~Wen, L.~Zhou,
  L.~Liu, C.~Guo, H.~Chen, S.~Yu, L.~Liu, and X.~Cai, ``{High-performance
  hybrid silicon and lithium niobate Mach--Zehnder modulators for 100 Gbit s -1
  and beyond},'' \emph{Nature Photonics}, vol.~13, no.~5, pp. 359--364, 2019.
  [Online]. Available: \url{http://dx.doi.org/10.1038/s41566-019-0378-6}
\BIBentrySTDinterwordspacing

\bibitem{Chen2011}
\BIBentryALTinterwordspacing
H.-W. Chen, J.~D. Peters, and J.~E. Bowers, ``{Forty Gb/s hybrid silicon
  Mach-Zehnder modulator with low chirp},'' \emph{Optics Express}, vol.~19,
  no.~2, p. 1455, jan 2011. [Online]. Available:
  \url{https://www.osapublishing.org/oe/abstract.cfm?uri=oe-19-2-1455}
\BIBentrySTDinterwordspacing

\bibitem{Hiraki2017}
\BIBentryALTinterwordspacing
T.~Hiraki, T.~Aihara, K.~Hasebe, K.~Takeda, T.~Fujii, T.~Kakitsuka,
  T.~Tsuchizawa, H.~Fukuda, and S.~Matsuo, ``{Heterogeneously integrated
  III-V/Si MOS capacitor Mach-Zehnder modulator},'' \emph{Nature Photonics},
  vol.~11, no.~8, pp. 482--485, 2017. [Online]. Available:
  \url{http://dx.doi.org/10.1038/nphoton.2017.120}
\BIBentrySTDinterwordspacing

\bibitem{Amin2018}
\BIBentryALTinterwordspacing
R.~Amin, R.~Maiti, C.~Carfano, Z.~Ma, M.~H. Tahersima, Y.~Lilach, D.~Ratnayake,
  H.~Dalir, and V.~J. Sorger, ``{0.52 V-mm ITO-based Mach-Zehnder Modulator in
  Silicon Photonics},'' \emph{APL Photonics}, vol.~3, no.~12, p. 126104, aug
  2018. [Online]. Available: \url{http://dx.doi.org/10.1063/1.5052635
  http://aip.scitation.org/doi/10.1063/1.5052635
  http://arxiv.org/abs/1809.03544}
\BIBentrySTDinterwordspacing

\bibitem{Green2007}
\BIBentryALTinterwordspacing
W.~M. Green, M.~J. Rooks, L.~Sekaric, and Y.~A. Vlasov, ``{Ultra-compact, low
  RF power, 10 Gb/s silicon Mach-Zehnder modulator},'' \emph{Optics Express},
  vol.~15, no.~25, p. 17106, 2007. [Online]. Available:
  \url{https://www.osapublishing.org/oe/abstract.cfm?uri=oe-15-25-17106}
\BIBentrySTDinterwordspacing

\bibitem{Rios2015}
\BIBentryALTinterwordspacing
C.~R{\'{i}}os, M.~Stegmaier, P.~Hosseini, D.~Wang, T.~Scherer, C.~D. Wright,
  H.~Bhaskaran, and W.~H.~P. Pernice, ``{Integrated all-photonic non-volatile
  multi-level memory},'' \emph{Nature Photonics}, vol.~9, no.~11, pp. 725--732,
  nov 2015. [Online]. Available:
  \url{http://www.nature.com/articles/nphoton.2015.182}
\BIBentrySTDinterwordspacing

\bibitem{milosevic_ion_2018}
\BIBentryALTinterwordspacing
M.~M. Milosevic, X.~Chen, W.~Cao, A.~F.~J. Runge, Y.~Franz, C.~G. Littlejohns,
  S.~Mailis, A.~C. Peacock, D.~J. Thomson, and G.~T. Reed,
  ``\BIBforeignlanguage{en}{Ion {Implantation} in {Silicon} for {Trimming} the
  {Operating} {Wavelength} of {Ring} {Resonators}},''
  \emph{\BIBforeignlanguage{en}{IEEE Journal of Selected Topics in Quantum
  Electronics}}, vol.~24, no.~4, pp. 1--7, Jul. 2018, 35 citations (Crossref)
  [2021-12-02]. [Online]. Available:
  \url{https://ieeexplore.ieee.org/document/8276286/}
\BIBentrySTDinterwordspacing

\bibitem{Feldmann:19}
\BIBentryALTinterwordspacing
J.~Feldmann, N.~Youngblood, C.~D. Wright, H.~Bhaskaran, and W.~H.~P. Pernice,
  ``All-optical spiking neurosynaptic networks with self-learning
  capabilities,'' \emph{Nature}, vol. 569, no. 7755, pp. 208--214, 2019.
  [Online]. Available: \url{https://doi.org/10.1038/s41586-019-1157-8}
\BIBentrySTDinterwordspacing

\bibitem{Gholipour:2015}
\BIBentryALTinterwordspacing
B.~Gholipour, P.~Bastock, C.~Craig, K.~Khan, D.~Hewak, and C.~Soci, ``Amorphous
  metal-sulphide microfibers enable photonic synapses for brain-like
  computing,'' \emph{Advanced Optical Materials}, vol.~3, no.~5, pp. 635--641,
  2015. [Online]. Available:
  \url{https://onlinelibrary.wiley.com/doi/abs/10.1002/adom.201400472}
\BIBentrySTDinterwordspacing

\bibitem{Chrostowski2014}
\BIBentryALTinterwordspacing
L.~Chrostowski, X.~Wang, J.~Flueckiger, Y.~Wu, Y.~Wang, and S.~T. Fard,
  ``{Impact of Fabrication Non-Uniformity on Chip-Scale Silicon Photonic
  Integrated Circuits},'' in \emph{Optical Fiber Communication
  Conference}.\hskip 1em plus 0.5em minus 0.4em\relax Washington, D.C.: OSA,
  2014, p. Th2A.37. [Online]. Available:
  \url{https://www.osapublishing.org/abstract.cfm?URI=OFC-2014-Th2A.37}
\BIBentrySTDinterwordspacing

\bibitem{Saeedi2015}
S.~Saeedi and A.~Emami, ``Silicon-photonic {PTAT} temperature sensor for
  micro-ring resonator thermal stabilization,'' \emph{Optics Express}, vol.~23,
  no.~17, p. 21875, 2015, 14 citations (Crossref) [2021-10-20].

\bibitem{huang_demonstration_2020}
\BIBentryALTinterwordspacing
C.~Huang, S.~Bilodeau, T.~Ferreira~de Lima, A.~N. Tait, P.~Y. Ma, E.~C. Blow,
  A.~Jha, H.-T. Peng, B.~J. Shastri, and P.~R. Prucnal,
  ``\BIBforeignlanguage{en}{Demonstration of scalable microring weight bank
  control for large-scale photonic integrated circuits},''
  \emph{\BIBforeignlanguage{en}{APL Photonics}}, vol.~5, no.~4, p. 040803, Apr.
  2020. [Online]. Available:
  \url{http://aip.scitation.org/doi/10.1063/1.5144121}
\BIBentrySTDinterwordspacing

\end{thebibliography}
\end{document}